\documentclass[twocolumn]{IEEEtran}

\usepackage[mathscr]{eucal}
\usepackage[cmex10]{amsmath}
\usepackage{epsfig,epsf,psfrag}
\usepackage{amssymb,amsmath,amsthm,amsfonts,latexsym}
\usepackage{amsmath,graphicx,bm,xcolor,url}
\usepackage[caption=false]{subfig} 
\usepackage{fixltx2e}%ordering of single and double column floats
\usepackage{array}%array and tabular environments
\usepackage{verbatim}
\usepackage{bm}
\usepackage{algorithmic, cite}
\usepackage{algorithm}
\usepackage{verbatim}
\usepackage{textcomp}
\usepackage{mathrsfs}
\usepackage{multirow}
\usepackage{epstopdf}
\catcode`~=11 \def\UrlSpecials{\do\~{\kern -.15em\lower .7ex\hbox{~}\kern .04em}} \catcode`~=13 

\allowdisplaybreaks[1]
 
\newcommand{\nn}{\nonumber}

\newcommand{\calC}{\mathcal{C}}

\newcommand{\calF}{\mathcal{F}}

\newcommand{\calN}{\mathcal{N}}

\newcommand{\calU}{\mathcal{U}}

\newcommand{\calX}{\mathcal{X}}
\newcommand{\calY}{\mathcal{Y}}

\newcommand{\rmb}{\mathrm{b}}

\newcommand{\rmd}{\mathrm{d}}

\newcommand{\bbE}{\mathbb{E}}

\newcommand{\bbN}{\mathbb{N}}

\newcommand{\bbP}{\mathbb{P}}

\newcommand{\bbR}{\mathbb{R}}

\DeclareMathAlphabet{\mathbsf}{OT1}{cmss}{bx}{n}
\DeclareMathAlphabet{\mathssf}{OT1}{cmss}{m}{sl}% slanted sans serif

\newcommand{\rvC}{\mathsf{C}}

\newcommand{\rvT}{\mathsf{T}}

\DeclareSymbolFont{bsfletters}{OT1}{cmss}{bx}{n}  
\DeclareSymbolFont{ssfletters}{OT1}{cmss}{m}{n}
\DeclareMathSymbol{\bsfGamma}{0}{bsfletters}{'000}
\DeclareMathSymbol{\ssfGamma}{0}{ssfletters}{'000}
\DeclareMathSymbol{\bsfDelta}{0}{bsfletters}{'001}
\DeclareMathSymbol{\ssfDelta}{0}{ssfletters}{'001}
\DeclareMathSymbol{\bsfTheta}{0}{bsfletters}{'002}
\DeclareMathSymbol{\ssfTheta}{0}{ssfletters}{'002}
\DeclareMathSymbol{\bsfLambda}{0}{bsfletters}{'003}
\DeclareMathSymbol{\ssfLambda}{0}{ssfletters}{'003}
\DeclareMathSymbol{\bsfXi}{0}{bsfletters}{'004}
\DeclareMathSymbol{\ssfXi}{0}{ssfletters}{'004}
\DeclareMathSymbol{\bsfPi}{0}{bsfletters}{'005}
\DeclareMathSymbol{\ssfPi}{0}{ssfletters}{'005}
\DeclareMathSymbol{\bsfSigma}{0}{bsfletters}{'006}
\DeclareMathSymbol{\ssfSigma}{0}{ssfletters}{'006}
\DeclareMathSymbol{\bsfUpsilon}{0}{bsfletters}{'007}
\DeclareMathSymbol{\ssfUpsilon}{0}{ssfletters}{'007}
\DeclareMathSymbol{\bsfPhi}{0}{bsfletters}{'010}
\DeclareMathSymbol{\ssfPhi}{0}{ssfletters}{'010}
\DeclareMathSymbol{\bsfPsi}{0}{bsfletters}{'011}
\DeclareMathSymbol{\ssfPsi}{0}{ssfletters}{'011}
\DeclareMathSymbol{\bsfOmega}{0}{bsfletters}{'012}
\DeclareMathSymbol{\ssfOmega}{0}{ssfletters}{'012}

\newcommand{\hatf}{\hat{f}}

\newcommand{\hatX}{\hat{X}}

\newcommand{\tilX}{\tilde{X}}

\newcommand{\hatY}{\hat{Y}}

\newcommand{\eps}{\varepsilon}
\DeclareMathOperator*{\argmin}{arg\,min}

\DeclareMathOperator{\var}{\mathrm{Var}}

\DeclareMathOperator{\cov}{\mathsf{Cov}}

\newtheorem{theorem}{Theorem} 
\newtheorem{lemma}{Lemma}

\newtheorem{corollary}{Corollary}
\newtheorem{definition}{Definition}

\newcommand{\qednew}{\nobreak \ifvmode \relax \else
      \ifdim\lastskip<1.5em \hskip-\lastskip
      \hskip1.5em plus0em minus0.5em \fi \nobreak
      \vrule height0.75em width0.5em depth0.25em\fi}

\usepackage{flushend}
\usepackage{dsfont}
\usepackage{xspace}
\usepackage{cite}
\allowdisplaybreaks[2]
\newcommand{\cR}{L}
\newcommand{\cS}{F}
\newcommand{\cD}{G}

\begin{document} 

%\title{On Gaussian Channels with Variable-Length Feedback and Non-Vanishing Error Probabilities} 
%\title{On AWGN Channels and Gaussian MACs with Variable-Length Feedback and Non-Vanishing Error~Probabilities} 
\title{On Gaussian MACs with  Variable-Length Feedback and Non-Vanishing Error~Probabilities} 
%\author{\IEEEauthorblockN{Lan V. Truong, and Vincent Y.~F.~Tan}\\
%\IEEEauthorblockA{Department of Electrical and Computer Engineering\\
%National University of Singapore \\ Emails: \url{lantruong@u.nus.edu}; \url{vtan@nus.edu.sg}}}

\author{Lan V.\ Truong, {\em Member, IEEE},  $\quad$ 
        Vincent Y.~F.~Tan, {\em Senior Member, IEEE}  \thanks{The authors are supported by an NUS Young Investigator Award (R-263-000-B37-133) and a  Singapore Ministry of Education (MOE) Tier 2 grant (R-263-000-B61-112). This paper was presented in part at the 2017 International Symposium on Information Theory.}
\thanks{L.\ Truong is with the Department of Electrical and Computer
Engineering, National University of Singapore, Singapore 117583 (e-mail:
lantruong@u.nus.edu).} \thanks{V.\ Y.\ F.\ Tan is with the the Department of Electrical and Computer
Engineering, National University of Singapore, Singapore 117583, and also
with the Department of Mathematics, National University of Singapore,
Singapore 119076 (e-mail: vtan@nus.edu.sg).}  \thanks{Communicated by A.\ Tchamkerten, Associate Editor for Shannon Theory.}
 \thanks{Copyright (c) 2017 IEEE. Personal use of this material is permitted.  However, permission to use this material for any other purposes must be obtained from the IEEE by sending a request to pubs-permissions@ieee.org.}}
%%%%%%%%%%%%%%%%%%%%%%%%%%%%%%%%%%%%%%%%%%%%%%%%%%%%%%%%%%%%%%
%%%%%%%%%%%%%%%%%%%%%%%%%%%%%%%%%%%%%%%%%%%%%%%%%%%%%%%%%%%%%%

\maketitle
 
\begin{abstract}  
We characterize the fundamental limits of transmission of information over a Gaussian multiple access channel (MAC) with   the use of variable-length feedback codes and under a non-vanishing error probability formalism. % For the AWGN channel, we establish the $\varepsilon$-capacity (for $0<\varepsilon<1$) and show that it is larger than the corresponding $\varepsilon$-capacity when fixed-length feedback is available.
We develop new achievability and converse techniques to handle the continuous nature of the channel and the presence of expected power constraints. %W show that a variable-length feedback with termination (VLFT) code outperforms a stop-feedback code  in terms of the second-order asymptotic behavior.  
  We establish the $\varepsilon$-capacity regions and bounds on the second-order asymptotics of the Gaussian MAC with variable-length feedback with termination (VLFT) codes and stop-feedback codes. We show that the former outperforms the latter significantly. Due to the multi-terminal nature of the channel model, we leverage  tools from renewal theory developed by  Lai and Siegmund  to bound the asymptotic behavior of the maximum of a finite number of  stopping times.  %  We do so by leveraging  tools from renewal theory developed by  Lai and Siegmund.
% to prove this bound and hence to %, we make use of   tools from 
%%To establish the bound,   we   develop new results  based on work on 
%%, to 
%establish the achievability result.
\end{abstract}   
\section{Introduction}
\subsection{Background and Related Works}\label{sub:1Arev}

Shannon~\cite{Sha56} showed that noiseless   feedback does not increase the capacity of point-to-point memoryless channels. Despite this seemingly negative result, it is known that feedback   significantly simplifies coding schemes and decreases the error probability. For example, Schalkwijk and  Kailath (SK)~\cite{SK66} proposed a simple coding scheme for the additive white Gaussian noise (AWGN) channel with fixed-length  feedback based on the idea of refining the receiver's knowledge of the   noise in each  transmission. The sender then iteratively corrects each  error  in the previous transmission. The error probability for this scheme is known to decay doubly exponentially fast in the blocklength.  Burnashev and Yamamoto~\cite{Burn14} showed that even  with noisy feedback, the reliability function of an AWGN channel improves (over the no feedback case). Ozarow~\cite{Ozarow} extended SK's coding scheme \cite{SK66} and showed that the capacity region of the Gaussian MAC is enlarged in the presence of feedback.  These ideas   are collectively known as {\em posterior matching}~\cite{ShayevitzF}. These ideas have  also  been extended by Truong, Fong and Tan~\cite{TruongFongTan17} to the case where the error probability is not required to vanish.

%However, Shannon also showed that it  may increase the zero-error capacity. 

It is also well known that  feedback can increase the capacity of channels with memory. Cover and Pombra~\cite{CoverPombra1989} characterized the feedback capacity of non-stationary additive  Gaussian noise channels with memory. Kim~\cite{YHK2010} found the capacity of the first-order autoregressive moving-average AWGN channel with feedback. For finite alphabet channels with memory and feedback, expressions of feedback capacity have been  derived for the trapdoor channel~\cite{HaimPermuter2008} and the Ising channel~\cite{HaimPermuter2014}. %In addition, feedback can help to reduce the encoding-decoding complexity and increase reliability (error exponent).  
It is also known that feedback can increase the second-order coding rates of certain discrete memoryless channels (DMCs)~\cite{AW14}. 
% For multiuser channels, it was mentioned that  feedback can enlarge the capacity region  of the Gaussian MAC~\cite{Ozarow}. It can also enlarge the capacity region of  memoryless MAC~\cite{GaarderWolf1975},~\cite{CoverLeung1981},~\cite{Ozarow},\cite{VenkataramananPradhan2011} and broadcast~\cite{Marton1979},~\cite{Ozarow1984},~\cite{VenkataramananPradhan2013}. Ozarow~\cite{Ozarow} found the capacity of the AWGN-MAC with feedback which strictly larger than the AWGN-MAC without feedback.
%For the Gaussian MAC Truong-Fong-Tan~\cite{TruongFongTan2015} found the $\eps$-capacity of the AWGN-MAC with feedback. 

%Nevertheless, it is known that feedback can be very useful provided that variable-length codes allowed. 
A greater advantage of feedback can be observed if one allows the length of the feedback signal to vary based on the quality of the channel output. Burnashev~\cite{Burnashev1976} demonstrated that the error exponent improves dramatically in this variable-length feedback  setting. In fact, the error exponent of a DMC with variable-length feedback is $E(R)= {C_1} \big(1-\frac{R}{C}\big)$
%\begin{align}
%E(R)= {C_1} \Big(1-\frac{R}{C}\Big)\label{eqn:burn}
%\end{align}
 for all rates $0\le R\le C$, where $C$ is the capacity of the DMC and $C_1$ is the maximal relative entropy between the conditional output distributions. Yamamoto and Itoh~\cite{YamamotoItoh1979} proposed a simple and conceptually important two-phase coding scheme that attains $E(R)$. While the error exponent results in \cite{Burnashev1976} and \cite{YamamotoItoh1979} are of paramount importance in feedback communications,  we focus on  the scenario in which the error probability is non-vanishing~\cite{TanBook}.  % Musy~\cite{Musy}  considered variable-length coding for  the MAC in which the decoder is allowed to decode the independent messages sent by the transmitters at different random times. 

For variable-length codes under the {\em non-vanishing error probability} formalism, Polyanskiy, Poor and  Verd\'u~\cite{Yury2011} provided non-asymptotic achievability and converse bounds for the coding rates. They also derived asymptotic expansions for the optimal code lengths of DMCs and showed dramatic improvements over the no feedback and the fixed-length feedback settings. In particular the channel dispersion vanishes, and so the backoff  from capacity  at finite blocklengths is significantly reduced. Trillingsgaard and Popovski~\cite{Trillingsgaard:2014} generalized the results for DMCs in~\cite{Yury2011} to the discrete memoryless  multiple access channel (DM-MAC). In it, they used ideas contained in Tan and Kosut \cite{TK14} and MolavianJazi and Laneman~\cite{Mol12} to analyze achievable second-order asymptotics for the DM-MAC. However, only achievability results were provided. It was also shown numerically in~\cite{Trillingsgaard:2014} that variable-length feedback outperforms fixed-length feedback.  % They showed that for BSC channel with capacity $1/2$ the blocklength required to achieve $90\%$ of the capacity os smaller than $200$, compared to at least $3100$ for the best fixed-blocklength code (even with noiseless feedback). 
Achievability and converse bounds under variable-length full-feedback (VLF) and variable-length stop-feedback (VLSF) for the binary erasure channel (BEC) have recently been derived by Devassy {\em et al.}~\cite{Devassy2016}. In addition, Trillingsgaard {\em et al.} used ideas related to the compound channel   \cite{Pol13b} to study the $2$-user~\cite{Trillingsgaard15} and  $K$-user~\cite{Trillingsgaard2016}  common-message discrete memoryless broadcast channel with stop-feedback.
%\begin{align}
%\frac{\rvC N}{1-\eps} -\sqrt{\frac{VN}{1-\eps}}\calE_a +o(\sqrt{N}) \leq \log M_{sf}^*(N,\eps) \leq \frac{\rvC N}{1-\eps} -\sqrt{\frac{VN}{1-\eps}}\calE_c +o(\sqrt{N}).
%\end{align} for some constants $\calE_a, \calE_c,V$. 
However, the techniques used in both the achievability and converse parts in \cite{Devassy2016}, \cite{Trillingsgaard15} and \cite{Trillingsgaard2016} are difficult to extend to Gaussian channels. This is because the authors leveraged the fact that a set of information densities for discrete channels can be bounded. This, together with  Hoeffding's inequality, allows the authors  to control the expectation of the maximum of a set of stopping times to eventually upper  bound   the average transmission time. The information density terms for Gaussian channels are not bounded. Hence, to study this important class of channels under variable-length feedback, we develop new techniques. We mention here that while the analysis of variable-length codes for non-vanishing error probabilities has been restricted to the finite alphabet setting, for the {\em vanishing error probability} formalism,    however, general alphabets have been considered both with and without cost constraints in the important works of Burnashev~\cite{Burnashev80} and Nakibo\u{g}lu and Gallager~\cite{Nak08}.

We characterize the information-theoretic  limits of the Gaussian MAC  when variable-length feedback is available at the encoder and a non-vanishing error probability is permitted. In particular, we circumvent the problem of the continuous nature of the alphabets by deriving new bounds on the moments (e.g., expectation and variance) of the maximum of a set of random variables (e.g., stopping times). These techniques may be of independent interest in other problems.

%in the converse proof~\cite[p.25]{Trillingsgaard2016} and achievability proof~\cite[p.43]{Trillingsgaard2016}. This require the information density functions to be bounded. More specifically, for the achievability proof, to bound $\bbE(\max_{k} \tau_k)$, they need to use the Hoeffding's inequality in~\cite[Equation (383)]{Trillingsgaard2016}, which requires the information density functions $i_{P_{X_n},W_k}(X_n;Y_{k,n})$ to be bounded. For the converse proof, to apply the Hoeffding's inequality in ~\cite[Equation~(182)]{Trillingsgaard2016}, $\tilde{i}_k(x,y), k=1,2,\ldots,K$  need to be bounded.  Note that this requirement is not satisfied in AWGN channels. 

\subsection{Main Contributions}\label{sec:main_contr}
We   propose a variable-length feedback model for Gaussian channels. We carefully define the expected power constraint  so that it is  analogous to the definition in the fixed-length feedback setting. In the latter setting, the   power constraint of a code for a point-to-point channel with (deterministic) blocklength $N\in\bbN$ is defined to be 
\begin{align}
 \bbE\bigg[\sum_{n=1}^N X_n^2\bigg]\leq NP,\label{eqn:exp_pow1}
\end{align}
where $X_n $ is the input to the channel at the $n$-th time slot and $P >0$ is the admissible power.  
However, in the variable-length feedback setting, the analogue of $N$, usually denoted as $\tau\in\bbN$, is a stopping  time (i.e., the random decoding time instant). Hence, one needs to carefully  define the analogue of \eqref{eqn:exp_pow1} so that we can utilize existing mathematical techniques for analyzing stopping times. We note  that the expected power constraint  we propose  in \eqref{eqn:mac_pow} is analogous to that in~\cite[Sec.~II.A]{Nak08}, i.e.,
\begin{equation}
\bbE\bigg[\sum_{n=1}^\tau X_n^2 \bigg]\le\bbE (\tau ) P.
\end{equation}
However, our formulation in \eqref{eqn:mac_pow} is  somewhat more convenient to analyze under the non-vanishing error probability formalism.%  in both the direct and converse parts.

In our main contribution, we derive achievability and converse bounds for the Gaussian MAC with two forms of variable-length feedback---stop-feedback and variable-length feedback with termination (VLFT).  We establish the $\eps$-capacity regions. We show that under the VLFT setting, we can achieve a larger $\eps$-capacity region compared to the stop-feedback setting.  We also provide bounds on the second-order terms. Our achievability proof for the Gaussian MAC with stop-feedback uses some non-standard techniques. We find  that Doob's optional stopping theorem~\cite[Thm.~10.10]{Williams}, which was used in~\cite{Yury2011} for the DMC, is not sufficient to bound the expected blocklength of the code. We develop new results, coupled with  work on renewal theory by Gut~\cite{Gut1974} and Lai and Siegmund~\cite{LaiSiegmund1979},  to bound the expected blocklength. The converse proof for the Gaussian MAC borrows some ideas from the weak converse proof in Ozarow's analysis for the Gaussian MAC with fixed-length feedback~\cite{Ozarow}. However,  our choice of  parameters is   different from~\cite{Ozarow}. This is  to account for the variable-length setting that we study. 

\subsection{Paper Organization}
The rest of this paper is structured as follows: In Section \ref{sec:awgn_mac}, we provide a precise problem setting for the Gaussian MAC, state the main results, and provide intuitions for these results. We also explain the novelties of our arguments relative to existing works. The achievability and converse proofs are provided in Sections \ref{sec:ach} and \ref{sec:conv} respectively. We conclude our discussion and suggest avenues for future work in Section~\ref{sec:conclu}. Auxiliary technical results that are not essential to the main arguments are relegated to the appendices.
\section{Gaussian MAC with Variable-Length Feedback}\label{sec:awgn_mac}
\subsection{Notation, Channel Model and Definitions}
\subsubsection{Notation} We use $\log x$ to denote the natural logarithm so information units throughout are in nats. We also define $x^+=\max(x,0)$ and $x^{-}=\max(-x,0)$. The Gaussian capacity and binary entropy functions are respectively defined as $\rvC(x) :=\frac{1}{2}\log(1+x)$ and $h_{\rmb}(x) :=-x\log x -(1-x) \log (1-x)$. 
The notation for random variables and information-theoretic quantities are standard and mainly follow the text by El Gamal and Kim~\cite{elgamal}. We use $\sigma(A)$ to denote the smallest $\sigma$-field on which random variable $A$ is measurable.  We write $\calN(\mu,\nu)$ for the univariate Gaussian distribution with mean $\mu$ and variance $\nu$. We also use standard asymptotic notation such as $O(\cdot)$.
\subsubsection{Channel Model}
The channel model is given by
\begin{align}
Y=X_1+X_2 +Z,
\end{align} where $X_1$ and $X_2$ represent the inputs to the channel, $Z \sim \mathcal{N}(0,1)$ is additive Gaussian noise with  zero mean and unit variance, and $Y$ is the output of the channel. Thus, the channel law from $(X_1,X_2)$ to $Y$ can be written as
\begin{align}
\bbP(y|x_1,x_2)=\frac{1}{\sqrt{2\pi}}\exp\left(-\frac{1}{2}(y-x_1-x_2)^2\right).\label{eqn:Wchannel}
\end{align}
%where $x_1, x_2,y\in\bbR$.

\subsubsection{Basic Definitions}
 The following definitions generalize~\cite{Yury2011} to the Gaussian MAC with expected power constraints.
\begin{definition}
\label{def1_mac}
An $(M_1,M_2,N,P_1,P_2,\eps)$ stop-feedback code for  the Gaussian MAC $\bbP(y|x_1,x_2)$, where $N,P_1,P_2$ are positive   numbers, $M_1,M_2$ are positive integers, and $0\leq \eps\leq 1$, is defined by:
\begin{enumerate}
\item Two spaces $\calU_1, \calU_2$ and probability distributions $P_{U_1}, P_{U_2}$ on them, defining independent random variables $U_j, j = 1,2$ each of which is revealed to transmitter $j = 1,2$ and the receiver  before the start of transmission; i.e., $(U_1,U_2)$ acts as common randomness.\footnote{The  common randomness is  used to initialize the encoders and the decoder before the start of transmission. See the usage of the Bernoulli random variable $D$ in Lemmas~\ref{them6} and~\ref{lem:vlft}. The reader is referred to  the analogue of this common randomness and accompanying discussions for the point-to-point case in~\cite{Yury2011}. }%, and $U_1$ is independent of $U_2$.
\item Two sequences of encoders $f_n^{(1)} :\calU_1 \times \{1,2,\ldots, M_1\} \to \bbR$ and $f_n^{(2)}:\calU_2 \times \{1,2,\ldots,M_2\} \to \bbR$ (indexed by $n\in\bbN$) defining channel inputs
$X_{jn}=f_n^{(j)}(U_j,W_j)$. 
where $W_j$ is equiprobable on the message set $\{1,2,\ldots,M_j\}$ for $j=1,2$.
\item A sequence of decoders $g_n: \calU_1\times \calU_2 \times \bbR^n \to \{1,2,\ldots,M_1\}\times \{1,2,\ldots,M_2\}$ providing estimates  $(W_1,W_2)$ at   times~$n$.
\item A non-negative integer-valued random variable $\tau$, a stopping time of the filtration $\{\sigma(U_1,U_2,Y^n)\}_{n=1}^{\infty}$, which satisfies
\begin{align}
\bbE (\tau) \leq N.
\end{align} 
\item The expected power constraints at the encoders
\begin{align}
\sum_{n=1}^\infty \bbE[X_{jn}^2] \leq \bbE(\tau) P_j, \quad j=1,2. \label{eqn:mac_pow}
\end{align}
\end{enumerate}
The final decision $(\hat{W}_1, \hat{W}_2)=g_{\tau}(U_1,U_2,Y^{\tau})$ is computed at time $\tau$ 
%as follows
%\begin{align}
%(\hat{W}_1, \hat{W}_2)=g_{\tau}(U_1,U_2,Y^{\tau}),
%\end{align}
and must satisfy
\begin{align}
\bbP[(\hat{W}_1, \hat{W}_2)\neq (W_1, W_2)] \leq \eps. \label{eqn:prob_mac}
\end{align}
\end{definition}
\begin{definition}
\label{def2_mac}
An $(M_1,M_2,N,P_1,P_2,\eps)$ variable-length feedback with termination code  (VLFT)  is defined as in Definition~\ref{def1_mac} except that $\tau$ is a stopping time of the filtration $\{\sigma(U_1,U_2,W_1,W_2,Y^n)\}_{n=1}^{\infty}$ and 
$
X_{jn}=f_n^{(j)}(U_j,W_j,Y^{n-1})$ for $j = 1,2$. 
\end{definition}
 
\subsection{Main Results and Discussions} \label{sec:main_res}
We now state our main results for the Gaussian MAC under various forms of variable-length codes with feedback. The proofs of the  achievability parts of Theorems \ref{stop_mac} and \ref{vlft_mac} are provided in Section  \ref{sec:ach}. The proofs of the  converse parts of Theorems~\ref{stop_mac} and \ref{vlft_mac} are provided in Section~\ref{sec:conv}. 
\begin{theorem}
\label{stop_mac}
For the Gaussian MAC $\bbP(y|x_1,x_2)$, there exists a sequence of $(M_1,M_2,N,P_1,P_2,\eps)$ stop-feedback codes
for any $(M_1,M_2)$ satisfying
\begin{align}
\label{eq166newq}
0 \leq \log M_j  &\leq \left(\frac{N}{1-\eps}-A\sqrt{\frac{N}{1-\eps}}  \right)\rvC(P_j) \nn\\*
&\qquad-\log N +O(1), \quad j=1,2\\
\label{eq167newq}
0 \leq \log M_1M_2& \leq \left(\frac{N}{1-\eps}-A\sqrt{\frac{N}{1-\eps}}  \right)\rvC(P_1+P_2)\nn\\*
&\qquad-\log N +O(1).
\end{align} where $A\ge 0$ is a constant given as
\begin{align}
\label{para_anewnew}
A&:=\min_{(i,j,k)\in \mathrm{perm}[3]  } \frac{1}{2} \Big(\sqrt{2(\cR_i+\cR_j)}+\sqrt{4\cR_k}\Big) \nn\\*
&\qquad+\frac{1}{4}\Big(\sqrt{2(\cR_i+\cR_k)}+ \sqrt{2(\cR_j+\cR_k)}\Big),
\end{align} and where  $\mathrm{perm}[3]$ is the set of all permutations of the tuple $(1,2,3)$ and
\begin{align}
\label{para_rnewnew}
\cR_j&:=\frac{4P_j}{(1+P_j)\left[\log(1+P_j)\right]^2},\quad j=1,2\\*
\label{para_rnewnew1}
\cR_3&:=\frac{4(P_1+P_2)}{(1+P_1+ P_2)\left[\log(1+P_1+P_2)\right]^2}.
\end{align}
Conversely, given any $(M_1,M_2,N, P_1,P_2,\eps)$ stop-feedback code, the following inequalities hold
\begin{align}
0 \leq \log M_j &\leq \frac{N\rvC(P_j)+h_{\rmb}(\eps)}{1-\eps},\quad j=1,2 \label{eqn:sf_up1}\\
0 \leq \log M_1M_2 &\leq \frac{N\rvC(P_1+P_2)+h_{\rmb}(\eps)}{1-\eps}.\label{eqn:sf_up3}
\end{align} 
\end{theorem}
\begin{theorem}
\label{vlft_mac}
Given a Gaussian MAC, for any $\rho \in [0,1]$, there exist a sequence of  $(M_1,M_2,N,P_1,P_2,\eps)$ VLFT-feedback codes for any $M_1, M_2$ satisfying 
\begin{align}
0\leq \log M_j &\leq  \frac{N\rvC(P_j(1-\rho^2))}{1-\eps}\nn\\*
&\qquad-\log \log N +O(1),\quad j=1,2\label{eqn:vlft_low1}\\
0\le\log M_1 M_2 &\leq  \frac{N\rvC(P_1+P_2+2\rho \sqrt{P_1P_2})}{1-\eps}\nn\\*
&\qquad-\log \log N+O(1).\label{eqn:vlft_low3}
\end{align}
Conversely, for any $(M_1,M_2,N, P_1,P_2,\eps)$-VLFT feedback code for the Gaussian MAC,   the following inequalities hold for some $\rho \in [0,1]$ and for $j=1,2$:
\begin{align}
0\leq \log M_j &\leq \frac{1}{1-\eps} \bigg[ N\rvC(P_j(1-\rho^2))+\nn\\*
&\qquad(N+1)h_{\rmb}\Big(\frac{1}{N+1}\Big)+h_{\rmb}(\eps) \bigg], \label{eqn:vlft_up1}\\
%&\leq \frac{N\rvC(P_j(1-\rho^2))+\log(N+1)+h_{\rmb}(\eps)+1}{1-\eps},\\
0\le\log M_1M_2 & \leq \frac{1}{1-\eps}  \bigg[N\rvC(P_1+P_2+2\rho \sqrt{P_1P_2}) \nn\\*
&\qquad+(N+1)h_{\rmb}\Big(\frac{1}{N+1}\Big)+h_{\rmb}(\eps) \bigg].\label{eqn:vlft_up3}
%&\leq \frac{N\rvC(P_1+P_2+2\rho \sqrt{P_1P_2})+\log(N+1)+h_{\rmb}(\eps)+1}{1-\eps}.
\end{align}
\end{theorem}

We define the $\eps$-capacity region  of a Gaussian MAC under the stop-feedback (resp.\ VLFT) formalisms  $\calC_{\mathrm{sf}}(P_1,P_2,\eps)$  (resp.\ $\calC_{\mathrm{t}}(P_1,P_2,\eps)$) to be the closure of the set of all rate pairs $(R_1,R_2)$ such that there exists a sequence of $(M_1,M_2,N,P_1,P_2,\eps)$ stop-feedback codes (resp.\ VLFT codes) such that 
$
\liminf_{N\to\infty}\frac{1}{N}\log M_j \ge R_j$ for $j=1,2$.
%,\quad j = 1,2 \label{eqn:liminf_rate}
%\end{equation}
and also that~\eqref{eqn:prob_mac} holds. Theorems~\ref{stop_mac} and~\ref{vlft_mac}  immediately imply the following corollary.% Similarly, we define the $\eps$-capacity region of an AWGN-MAC under the VLFT formalism $\calC_{\mathrm{t}}(P_1,P_2,\eps)$ to be the closure of the set of all rate pairs $(R_1,R_2)$ such that there exists a sequence of  $(M_1,M_2,N,P_1,P_2,\eps)$  VLFT-feedback codes such that  \eqref{eqn:liminf_rate} and \eqref{eqn:prob_mac} hold. 

\begin{corollary}\label{cor:eps_cap}
Let $0<\eps<1$. The $\eps$-capacity region $\calC_{\mathrm{sf}}(P_1,P_2,\eps)$ is the set of all $(R_1, R_2) \in\bbR_+^2$ satisfying 
\begin{align}
R_j &\le \frac{\rvC(P_j)}{1-\eps} ,\quad j=1,2   \label{eqn:cor1} \\
R_1+R_2 &\le\frac{\rvC(P_1+P_2)}{1-\eps}.
\end{align}
Similarly, the $\eps$-capacity region  $\calC_{\mathrm{t}}(P_1,P_2,\eps)$  is the set of all $(R_1, R_2) \in\bbR_+^2$ satisfying 
\begin{align}
R_j  & \le \frac{\rvC(P_j (1-\rho^2))}{1-\eps} ,\quad j=1,2 \label{eqn:eps_cap_stop1} \\ 
R_1 + R_2 &\le  \frac{\rvC(P_1+P_2 + 2\rho\sqrt{P_1 P_2} )}{1-\eps} \label{eqn:eps_cap_stop2}
\end{align}
for some $\rho\in [0,1]$. 
\end{corollary}
%\begin{remark}
%Besides the notes from AWGN channel, the AWGN-MAC has some additional notes
%\begin{itemize}
%\item K. F. Trillingsgaard and P. Popovski~\cite{Trillingsgaard:2014} generalized~\cite{Yury2011} to the DM-MAC. However, they could not bound $\bbE(\max_{k} \tau_k)$, hence only some numerical results were obtained to show that stop-feedback can increase the first-order coding rate ($\eps$-capacity). In our proof of achievability part for Theorem~\ref{stop_mac}, we showed a way to bound it for the AWGN-MAC. The arguments in Lemma~\ref{lem1b} are non-standard. 
%\item  The $\eps$-capacity region in Theorem~\ref{vlft_mac} is strictly larger than the one in Theorem~\ref{stop_mac}, which indicates more clearly the fact that full-feedback at encoders can help to increase the capacity region compared when this does not happen in variable-length settings. 
%\item For the converse proof for the VLFT code  in Theorem~\ref{vlft_mac}, we borrow some ideas from Ozarow converse proof~\cite{Ozarow}. However, the way we set parameters are slightly different from Ozarow (e.g.~\eqref{eq441newq} and~\eqref{eq442newq}). 
%\end{itemize}
%\end{remark}

Some remarks concerning Theorems \ref{stop_mac} and \ref{vlft_mac} and Corollary \ref{cor:eps_cap} are now in order:
\begin{enumerate}
%\item \textcolor{red} {Similar to the single-user case, for VLFT codes, the maximum delay is bounded by $N/(1-\eps)$ (cf.\ the proof of Theorem~\ref{vlft_mac}). This is an advantage of   VLFT codes compared with stop-feedback codes, where the maximum delay is unbounded.} 

\item  Trillingsgaard and Popovski~\cite{Trillingsgaard:2014} generalized the point-to-point variable-length feedback results for the DMC in Polyanskiy, Poor and Verd\'u~\cite{Yury2011} to the DM-MAC. In it, they used ideas contained in Tan and Kosut \cite{TK14} and MolavianJazi and Laneman \cite{Mol12} to analyze   achievable second-order asymptotics for the DM-MAC with variable-length feedback. However, Trillingsgaard and Popovski~\cite{Trillingsgaard:2014} could not analytically bound the expectation of the maximum of several   stopping times $\bbE(\max_{k} \tau_k)$ and they also could not prove a matching (first-order) converse. Instead, they provided numerical results to show that stop-feedback increases the first-order coding rate compared to the fixed-length feedback setting.

%In addition, variable-length codes with feedback also help to improve the first-order capacity since without variable-length codes, we only achieve $R_j\leq \rvC(P_j(1-\rho^2)/(1-\eps)), j=1,2$ and $ R_1+R_2 \leq \rvC((P_1+P_2+2\rho\sqrt{P_1P_2})/(1-\eps))$~\cite{TruongFongTan17} for fixed-length codes with feedback.

\item The multiplicative gains  of $\frac{1}{1-\eps}$ in \eqref{eqn:cor1}--\eqref{eqn:eps_cap_stop2} are due to the non-vanishing nature of the error probability and the use of variable-length codes with feedback. Note that for the Gaussian MAC without feedback, the strong converse holds in the sense that the $\eps$-capacity is independent of $\eps$~\cite{FongTan16}. 

\item %Inequalities \eqref{eqn:eps_cap_stop1}--\eqref{eqn:eps_cap_stop2} in Corollary \ref{cor:eps_cap} describe the $\eps$-capacity region using VLFT codes.
 The $\eps$-capacity region for VLFT codes is easily seen to be strictly larger than the corresponding region  for   fixed-length feedback codes recently studied  by Truong, Fong and Tan~\cite{TruongFongTan17}. In that scenario, the $\eps$-capacity region is given by~\cite{TruongFongTan17}  
\begin{align}
R_j &\leq \rvC \left( \frac{P_j(1-\rho^2)}{ 1-\eps} \right),\quad  j=1,2\\
  R_1+R_2 &\leq \rvC\left(\frac{P_1+P_2+2\rho\sqrt{P_1P_2}}{ 1-\eps}\right),
\end{align} 
for some $\rho\in [0, 1]$. 
 The enlargement is due to the following   consequence of Jensen's inequality:
\begin{equation}
\rvC \left( \frac{P}{1-\eps} \right) <\frac{\rvC(P)}{1-\eps} ,\qquad\forall\,  (P,\eps) \in (0,\infty)\times (0,1).
\end{equation}
% which follows from Jensen's inequality.
 This gain is present as variable-length feedback codes are  {\em adaptive}, i.e., their lengths are adapted to the quality of $Y^\infty$.
   %In our proof of the achievability part of Theorem~\ref{stop_mac}, we provide a simple method to upper bound the important quantity  $\bbE(\max_{k} \tau_k)$ for the Gaussian MAC, leading to  tight first-order terms in Corollary \ref{cor:eps_cap}. These non-standard arguments are mainly contained in Lemma~\ref{lem1b} to follow. We are unable to do the same for the DM-MAC due to a technical requirement of  Lemma \ref{lem0} (to follow) that requires relevant  information density random variables to be nonlattice. 

\item The $\eps$-capacity region  $\calC_{\mathrm{t}}(P_1,P_2,\eps)$ is strictly larger than $\calC_{\mathrm{sf}}(P_1,P_2,\eps)$, which  clearly illustrates the fact that feedback at encoders can enlarge the $\eps$-capacity region compared to the case where only stop-feedback is available. %In the  point-to-point scenario where it is shown in Corollary \ref{cor:eps_cap} that the $\eps$-capacities are the same under both settings but only the achievable second-order term using VLFT codes is superior to that using stop-feedback codes (cf.\ the lower bounds in~\eqref{eqn:stop_res} and \eqref{eqn:vlft1}).
 That $\calC_{\mathrm{t}}(P_1,P_2,\eps)$ is strictly larger than $\calC_{\mathrm{sf}}(P_1,P_2,\eps)$ is  completely analogous to the fact that fixed-length  feedback enlarges the capacity region of the Gaussian MAC (cf.\  Ozarow~\cite{Ozarow}).
 
 \item In the achievability proofs, we note that  Polyanskiy, Poor and Verd\'u~\cite{Yury2011} utilize the fact that the relevant information density random variable $i(X;Y)$  (induced by the capacity-achieving input distribution and the channel) is bounded when the channel is a  DMC~\cite[Eqn.~(107)]{Yury2011}. However, this fact does not hold for the AWGN channel and so our achievability proofs require  some  novel elements. All previous works on variable-length feedback for systems with non-vanishing error probabilities  \cite{Yury2011,Devassy2016,Trillingsgaard15,Trillingsgaard2016} involve channels with {\em discrete} alphabets.  In addition, we  leverage novel bounds (Lemmas \ref{lem0} and \ref{lem1b}) that control the first  and second moments of  the maximum of a set of  stopping times $\max_{k} \tau_k$ and multi-user information spectrum methods \cite{TK14,Mol12,Trillingsgaard:2014}.  %Stronger arguments for both the achievability and converse parts are needed to obtain a tight second-order term  in Theorem~\ref{stop_mac}. Note that in the point-to-point setting~\cite{Yury2011}, one can achieve zero dispersion (i.e., no $\sqrt{N}$ backoff term in \eqref{eq166newq} and \eqref{eq167newq}), so our results are unlikely to be second-order tight. We defer the derivation of tight second-order terms to future work but we note that our arguments are sufficient to obtain tight first-order terms (Corollary \ref{cor:eps_cap}). Such techniques were  hitherto unavailable. 

\item For the converse of Theorem~\ref{stop_mac}, we make use of   Fano-like arguments. Although some of the ideas are inspired by~\cite{Yury2011}, we need to augment   the original arguments so that the proof is amenable to Gaussian channels. More specifically, in~\cite{Yury2011}, the authors use the fact that the capacity of the DMC is $\sup \{I(\hat{X};\hat{Y}): P_{\hat{X}}(\rvT)=0\}$,    where $\rvT$ is a new symbol appended to the input and output alphabets of the DMC to form $\hat{\calX}$ and $\hat{\calY}$ respectively and $\hatX\in\hat{\calX}$ is the input random variable of the new DMC. However, for Gaussian MAC with the expected power constraints  in \eqref{eqn:mac_pow}, this does not hold.

\item For the converse proof of Theorem~\ref{vlft_mac}, we borrow some ideas from Ozarow's weak converse proof for the Gaussian MAC with   fixed-length feedback~\cite{Ozarow}. However, our parameter settings and the manipulations of  the resultant bounds are   different from Ozarow. See~\eqref{eq441newq} and~\eqref{eq442newq} in Lemma \ref{lem:conv_mac_vlft} to follow. 
 
\item Specializing our results for the Gaussian MAC to the   point-to-point  AWGN channel also yields novel results. In this case, 
 the first-order terms for VLFT code and stop-feedback codes are identical and equal to  to $\frac{\rvC(P)}{1-\eps}$; this can be seen by setting $\rho=0$ in Theorem~\ref{vlft_mac}. However, the achievability result for VLFT codes is better than the corresponding one for stop-feedback codes in the second-order term ($-O(\log\log N)$ compared to $-O(\sqrt{N})$). 
\end{enumerate}
\section{Achievability Proofs } \label{sec:ach}
\subsection{Achievability Proof   for Theorem~\ref{stop_mac}} \label{sec:ach_s}
To prove the achievability result for Theorem~\ref{stop_mac} in~\eqref{eq166newq} and~\eqref{eq167newq}, we  commence with some technical  results in Lemmas~\ref{lem00} to~\ref{lem1b}.  The achievability result for Theorem \ref{stop_mac} follows from a combination of Lemmas \ref{lem1} and \ref{them6} to follow. 
\begin{definition}[Strongly nonlattice \cite{Stone1965}] We say that a distribution function $F$ is strongly nonlattice if 
\label{stronglattice}
$
\liminf_{|t|\to \infty}|1-f(t)| >0,
$ where
$
f(t):=\int_{-\infty}^{\infty} e^{itx}\,\rmd F(x)
$ is the characteristic function of $F$. This is equivalent to Cramer's condition (C), i.e., that 
$
\limsup_{|t|\to \infty} |f(t)| < 1. 
$
\end{definition}
%The following lemma is adapted from Gut~\cite[Theorem~2.6]{Gut1974}.
\begin{lemma}[Asymptotics of Expected Values of Stopping Times]
\label{lem00}
Let $X_1,X_2,\ldots$ be i.i.d.\ random variables with positive mean $\mu=\bbE [X_1]$, finite variance $\sigma^2=\var(X_1)$ and $\bbE[X_1^+]<\infty$. Let $S_n:=X_1+X_2+\ldots+X_n$. For each $b\geq 0$ define
\begin{align}
\label{eq19newl}
\tau&=\tau(b)=\inf\{n: S_n>b\},\\*
\label{eq20newl}
\tau_+&=\tau(0)=\inf\{n:S_n>0\}.
\end{align}
Assume that $X_1$ has a distribution function $F_{X_1} $ that is strongly nonlattice in the sense of Definition \ref{stronglattice}. Then as $b\to \infty$,
\begin{align}
\mu \bbE (\tau )=b + \frac{\bbE (S_{\tau_+}^2)}{2\bbE (S_{\tau_+})}+o(1). 
\end{align}
\end{lemma}
\begin{IEEEproof}
Follows from Gut~\cite[Thm.~2.6]{Gut1974} and Wald's identity~\cite[Eqn.~(13) in Sec.~12.5]{Grimmett}.  %\cite[Problem 22.8(a)]{Billingsley}. 
%In the original statement of~\cite[Theorem 2.6]{Gut1974}, it is stated that
%\begin{align}
%\bbE (\tau) =\frac{b}{\mu} +\bbE(\tau_+)\frac{\bbE ( S_{\tau_+}^2)}{2(\bbE (S_{\tau_+}))^2}+o(1). \label{eqn:gut}
%\end{align}
%Now we claim that (i)  $\bbE (\tau_+)<\infty$   and (ii) $\bbE[|X_1| ] <\infty$.  Claim (i) follows by first noticing that the finiteness of $\mu=\bbE[ X_1]$ and  $\bbE [X_1^+]$ implies the same for $\bbE[X_1^-] =\bbE[X_1^+]-\bbE[X_1]$. By \cite[Theorem~2.1(a)]{Gut1974} (with $r=1$ in the statement therein), we have that  $\bbE (\tau_+)<\infty$. Claim (ii) follows because the finiteness of $\bbE[ X_1^-]$ and  $\bbE [X_1^+]$ implies the same for $\bbE[|X_1|] = \bbE[X_1^+] + \bbE[X_1^-] $. 
%
%%Due to the finiteness of $\mu=\bbE[ X_1]$ and  $\bbE [X_1^+]$, this means that $\bbE[X_1^-]<\infty$. Hence, by \cite[Theorem 2.1(a)]{Gut1974}, we have that  $\bbE (\tau_+)<\infty$  (take $r=1$ in the statement in \cite[Theorem 2.1(a)]{Gut1974}). Furthermore, $\bbE[ |X_{1}| ]< \infty $ also because $\mu <\infty$ and $\bbE[X_1^+]<\infty $. 
%
%Equipped with these two claims and~\cite[Equation (13) in Section 12.5]{Grimmett} (also see~\cite[Problem 22.8(a)]{Billingsley}),   Wald's identity
%\begin{align}
%\mu \bbE(\tau_+)=\bbE (S_{\tau_+} ) \label{eqn:wald}
%\end{align}
%applies. Uniting  \eqref{eqn:gut} and \eqref{eqn:wald}, we see that 
%%Therefore,  
%Lemma~\ref{lem00} follows.
\end{IEEEproof}
\begin{lemma}[Asymptotics of Variance of Stopping Times~\cite{LaiSiegmund1979}] \label{lem0}
Let $X_1,X_2,\ldots$ be i.i.d.\ random variables with positive mean $\mu$ and finite variance $\sigma^2$ and $\bbE(X_1^+)<\infty$. Let $S_n:=X_1+X_2+\ldots+X_n$. For each $b\geq 0$ define $\tau$ and $\tau_+$ as in~\eqref{eq19newl} and~\eqref{eq20newl}. If $X_1$ has a distribution function that  is strongly nonlattice, then as $b\to \infty$,
\begin{align}
\var (\tau) =\mu^{-3} \sigma^2 b + \mu^{-2} K +o(1),
\end{align}
where $K$ is a constant that does not depend on $b$ and is given by
\begin{align}
\label{eq38e}
K&:= \frac{\sigma^2 \bbE S_{\tau_+}^2}{2\mu \bbE S_{\tau_+}}+\frac{3}{4}\left(\frac{\bbE S_{\tau_+}^2}{\bbE S_{\tau_+}}\right)^2 -\frac{2}{3} \left(\frac{\bbE S_{\tau_+}^3}{\bbE S_{\tau_+}}\right) \nn\\*
&\qquad  - \left(\frac{\bbE S_{\tau_+}^2}{\bbE S_{\tau_+}}\right) \bbE\Big\{\min_{n\geq 0}S_n\Big\} \nn\\*
&\qquad-2\int_0^{\infty} \bbE\{S_{\tau(x)}-x\}\bbP\Big\{\min_{n\geq 0} S_n \leq -x\Big\} \,\rmd x.
\end{align}
\end{lemma}

%The following result can be regarded as a generalization of Wald's equation.} 
\begin{lemma}[Generalization of Wald's equation~\cite{wikiwald}\footnote{The proof of Lemma \ref{lem3a} in \cite{wikiwald} has been verified correct by the authors and the Associate Editor Prof.\ A.\ Tchamkerten~\cite{personalCommAE}. We thank the editor for his kind assistance. }]
\label{lem3a}
Let $\{X_n\}_{n=1}^{\infty}$ be an infinite sequence of real-valued random variables and let $\tau$ be a non-negative integer-valued random variable. Assume that
\begin{itemize}
\item $\{X_n\}_{n=1}^{\infty}$ are all integrable (finite-mean) random variables;
\item for all natural numbers $n$,
$\bbE[X_n 1\{\tau\geq n\}]=\bbE[X_n]\bbP(\tau \geq n)$;
\item the infinite series  
 $\sum_{n=1}^{\infty} \bbE[|X_n|1\{\tau\geq n\}]<\infty$;

\item $\{X_n\}_{n=1}^{\infty}$ all have the same expectation, and
\item $\tau$ has finite expectation.
\end{itemize}
Define $
S_{\tau}:=\sum_{n=1}^{\tau} X_n$. 
Then, we have
\begin{align}
\label{ward}
\bbE[S_{\tau}]=\bbE[\tau]\bbE[X_1].
\end{align}
\end{lemma}
Note that this is indeed a generalization of the standard Wald's equation~\cite{Bruss,Grimmett} which states that if $\{X_n\}_{n=1}^\infty$ is a sequence of i.i.d.\ integrable random variables and $\tau$ is a finite expectation stopping time with respect to $\{X_n\}_{n=1}^\infty$, then \eqref{ward} holds. Lemma \ref{lem3a} does not require $\{X_n\}_{n=1}^\infty$ to be i.i.d.\ The proof of Lemma \ref{lem3a}, which can be found in \cite{wikiwald}, is similar to that of Wald's equation~\cite{Bruss,Grimmett}.

%The fourth important technical lemma will eventually allow us to control the asymptotic behavior of the expectation of the maximum of several stopping times.
\begin{lemma}[Expectation of the Maximum of Random Variables]\label{lem1b}
Let $\{(X_{1N},X_{2N},X_{3N})\}_{N\ge 1}$ be three sequences of random variables satisfying %the following properties for some  {constants} $A\geq 0, D\geq 0$ and $B_1,B_2,B_3 \in \bbR$:
\begin{align}
\bbE [X_{jN} ]&= N-A\sqrt{N}-\cD-B_j+o(1),\quad j=1,2,3
\end{align}   for some  {constants} %$A\geq 0, \cD\geq 0$ and %
$B_1,B_2,B_3 \in \bbR$,  where  $A$ as given in~\eqref{para_anewnew} and $G$ is defined as follows:
\begin{align}
\cD&:=-\frac{1}{4}\left(B_{i_0}+B_{j_0}+2B_{k_0}\right) \nn\\*
&\qquad+\frac{1}{2}\left(\sqrt{2|\cS_{i_0}+\cS_{j_0}|+(B_{i_0}-B_{j_0})^2}\right)\nn\\*
&\qquad +\frac{1}{4}  \bigg(\sqrt{2|\cS_{i_0}+\cS_{k_0}|+(B_{i_0}-B_{k_0})^2} \nn\\*
&\qquad \qquad + \sqrt{2|\cS_{j_0}+\cS_{k_0}|+(B_{j_0}-B_{k_0})^2} \bigg),
\end{align} 
where 
\begin{align}
(i_0,j_0,k_0) &:=\argmin_{(i,j,k)\in \mathrm{perm}[3] } \frac{1}{2}  \sqrt{2(\cR_i+\cR_j)}   \nn\\*
&\quad+\frac{1}{4} \left(\sqrt{2(\cR_i+\cR_k)}+ \sqrt{2(\cR_j+\cR_k)} \right).
\end{align}
%we have
%\begin{align}
%  for some  {constants} %$A\geq 0, \cD\geq 0$ and %
%$B_1,B_2,B_3 \in \bbR$ and
Furthermore assume that 
\begin{align}
\var(X_{jN}) &\leq  \cR_j N+ \cS_j+o(1),\quad j=1,2,3
\end{align}
for some other constants $\cR_1>0,\cR_2>0,\cR_3>0$ and $\cS_1,\cS_2,\cS_3 \in \bbR$.  %Here $o(1)$ denotes an arbitrary sequence that vanishes as $N\to\infty$. 
%Then for $A$ as given in~\eqref{para_anewnew}, and 
%\begin{align}
%(i_0,j_0,k_0)&:=\argmin_{(i,j,k)\in \mathrm{perm}[3] } \frac{1}{2}  \sqrt{2(\cR_i+\cR_j)} +\frac{1}{4} \left(\sqrt{2(\cR_i+\cR_k)}+ \sqrt{2(\cR_j+\cR_k)} \right),\\
%\cD&:=-\frac{1}{4}\left(B_{i_0}+B_{j_0}+2B_{k_0}\right)+\frac{1}{2}\left(\sqrt{2|\cS_{i_0}+\cS_{j_0}|+(B_{i_0}-B_{j_0})^2}\right)\nn\\
%&\quad +\frac{1}{4}\left(\sqrt{2|\cS_{i_0}+\cS_{k_0}|+(B_{i_0}-B_{k_0})^2}+ \sqrt{2|\cS_{j_0}+\cS_{k_0}|+(B_{j_0}-B_{k_0})^2}\right),
%\end{align} 
Then, we have
\begin{align}
\bbE(\max\{X_{1N},X_{2N},X_{3N}\}) \leq N+o(1).
\end{align}
\end{lemma}

\begin{IEEEproof}
The proof is deferred to Appendix~\ref{app:prf}.
\end{IEEEproof}

\begin{lemma} \label{lem1} Consider a standard Gaussian MAC $\bbP(y|x_1,x_2)$ with expected power constraints $P_1,P_2$. For any $N'>0$, and $(M_1,M_2)$ satisfying
\begin{align}
\label{eq213newq}
0 \leq \log M_j &\leq (N'-A\sqrt{N'})\rvC(P_j)  \nn\\*
&\qquad -\log N'+O(1),\quad j=1,2\\
\label{eq215newq}
0 \leq \log M_1M_2 &\leq (N' -A\sqrt{N'})\rvC(P_1+P_2) \nn\\*
&\qquad-\log N'+O(1),
\end{align} we can find an $(M_1,M_2,N'+o(1),\frac{1}{N'})$ stop-feedback code with $A$ defined as in~\eqref{para_anewnew}.
\end{lemma}
\begin{IEEEproof}
Part of the proof is based on~\cite{Yury2011} and~\cite{Trillingsgaard:2014} but as mentioned, we need to combine existing ideas with Lemmas~\ref{lem0} and~\ref{lem1b} above. First, we show that there exists an $(M_1,M_2,N'+o(1),P_1,P_2,\frac{1}{N'})$ stop-feedback code with stopping time $\tau^*$, where
$
\bbE(\tau^*)  \leq N'+o(1),
$  the sizes of the message sets $M_1,M_2$ satisfy~\eqref{eq213newq} and~\eqref{eq215newq}, and finally,
$
\bbE\big[\sum_{n=1}^{\tau^*} X_{jn}^2\big]=\bbE(\tau^*) P_j,
$ for  $j=1,2$.
To define this code, we define two random variables $U_1$ and $ U_2$ each with distribution $\bbP_{U_j}:=(\bbP_{X_j})^{\infty}\times (\bbP_{X_j})^{\infty}\times \ldots \times (\bbP_{X_j})^{\infty}$  ($M_j$ times)
%\begin{align}
%\calU_j&:=\bbR^{\infty} \times \bbR^{\infty}\times \cdots \times \bbR^{\infty},\quad \mbox{($M_j$ times)}\\
%%\calU_2&:=\bbR^{\infty} \times \bbR^{\infty}\times \cdots \times \bbR_2^{\infty},\quad \mbox{($M_2$ times)}\\
%\bbP_{U_j}&:=(\bbP_{X_j})^{\infty}\times (\bbP_{X_j})^{\infty}\times \cdots \times (\bbP_{X_j})^{\infty}, \quad \mbox{($M_j$ times)},
%%\bbP_{U_2}&:=(\bbP_{X_2})^{\infty}\times (\bbP_{X_2})^{\infty}\times \cdots \times (\bbP_{X_2})^{\infty},\quad \mbox{($M_2$ times)},
%\end{align} 
where $j = 1,2$ and  $\bbP_{X_j} \sim \calN(0,P_j)$.

We generate the codebook as follows. For a realization of $U_1$, we generate $M_1$ i.i.d.\ infinite dimensional vectors $\{{\bf C}^{(1)}_j\}$ from $\bbP_{X_1} \sim \calN(0,P_1)$. Similarly, for each realization of $U_2$, we generate $M_2$ i.i.d.\ infinite dimensional vectors $\{{\bf C}^{(2)}_k\}$ from  $\bbP_{X_2} \sim \calN(0,P_2)$. The encoder and decoder depend on $U_1,U_2$ implicitly through $\{{\bf C}^{(1)}_j\}$ and $\{{\bf C}^{(2)}_k\}$. 

Encoder $j = 1,2$ consists of a sequence  of encoders $f_n^{(j)}$ that maps message  $w_j \in \{1,2,\ldots,M_j\}$ to an infinite sequence of inputs ${\bf C}_{w_j}^{(j)} \in \bbR^{\infty}$. %The encoding scheme 2 consists of sequences of encoders $f_n^{(2)}$ that maps messages $k \in \{1,2,\ldots,M_2\}$ to an infinite sequence of input ${\bf C}_k^{(2)} \in \bbR^{\infty}$. 
The mappings are without regard to feedback,
$
X_{jn}=f_n^{(j)}(w_j):={\bf C}^{(j)}_{w_j,n}$, 
%\end{align} 
where ${\bf C}^{(1)}_{w_1,n}$ and ${\bf C}^{(2)}_{w_2,n}$ are respectively the $n$-th coordinates of the infinite vectors ${\bf C}^{(1)}_{w_1}$ and ${\bf C}_{w_2}^{(2)}$.

Let  ${\bf C}^{(1)}_j(n):=(  {\bf C}^{(1)}_{j,1}  ,  \ldots , {\bf C}^{(1)}_{j,n})$ and similarly define ${\bf C}^{(2)}_k(n) $.  At   time $n$, the decoder computes the   (conditional) information densities:
\begin{align}
S_{j,k}^{(1,n)}:=i({\bf C}^{(1)}_j(n); Y^n|{\bf C}^{(2)}_k (n)),\\
S_{j,k}^{(2,n)}:=i({\bf C}^{(2)}_k(n); Y^n|{\bf C}^{(1)}_j (n)),\\
S_{j,k}^{(3,n)}:=i({\bf C}^{(1)}_j(n),{\bf C}^{(2)}_k(n);Y^n),
\end{align}
for all $(j,k) \in \{1,2,\ldots,M_1\}\times \{1,2,\ldots,M_2\}$,  
where
\begin{align}
\label{eq226newq}
&i({\bf C}^{(1)}_j(n); Y^n|{\bf C}^{(2)}_k(n)) \nn\\*
&:= \log \! \frac{\mathrm{d}\bbP_{X_1^n Y^n|X_2^n}}{\mathrm{d}(\bbP_{X_1^n|X_2^n} \! \times \! \bbP_{Y^n|X_2^n})}\left({\bf C}^{(1)}_j(n),{\bf C}^{(2)}_k(n),Y^n\right),
%\\
%i({\bf C}^{(2)}_k(n); Y^n|{\bf C}^{(1)}_j(n))&:=\log \frac{\mathrm{d}\bbP_{X_2^n Y^n|X_1^n}}{\mathrm{d}(\bbP_{X_2^n|X_1^n} \times \bbP_{Y^n|X_1^n})}\left({\bf C}^{(1)}_j(n),{\bf C}^{(2)}_k(n),Y^n\right),\\
%\label{eq228newq}
%i({\bf C}^{(1)}_j(n),{\bf C}^{(2)}_k(n);Y^n)&:=\log \frac{\mathrm{d}\bbP_{X_1^n X_2^n Y^n}}{\mathrm{d}(\bbP_{X_1^n X_2^n} \times \bbP_{Y^n})}\left({\bf C}^{(1)}_j(n),{\bf C}^{(2)}_k(n),Y^n\right).
\end{align}
and similarly for $i({\bf C}^{(2)}_k(n); Y^n|{\bf C}^{(1)}_j(n))$ and $i({\bf C}^{(1)}_j(n),{\bf C}^{(2)}_k(n);Y^n)$. 
For a triple of positive real numbers $(\gamma_1,\gamma_2,\gamma_3)$ to be   chosen later, the decoder also defines a number of stopping times as follows:
\begin{align}
\tau^{(1)}_{j,k}&:=\inf\{n\geq 0: i({\bf C}^{(1)}_j(n); Y^n|{\bf C}^{(2)}_k(n)) > \gamma_1\},\\
\tau^{(2)}_{j,k}&:=\inf\{n\geq 0: i({\bf C}^{(2)}_k(n); Y^n|{\bf C}^{(1)}_j(n)) > \gamma_2\},\\
\tau^{(3)}_{j,k}&:=\inf\{n\geq 0: i({\bf C}^{(1)}_j(n),{\bf C}^{(2)}_k(n);Y^n) > \gamma_3\},
\end{align}
and $\tau_{j,k}:=\max\{\tau^{(1)}_{j,k},\tau^{(2)}_{j,k},\tau^{(3)}_{j,k}\}$. 
The final decision is made by the decoder at the stopping time 
\begin{align}
\label{eq37a}
\tau^*:=\min_{j,k}\tau_{j,k}.
\end{align} 
The output of the decoder is given by
\begin{align} 
\label{eq:eq116rev}
g(Y^{\tau^*})=\max\{(j,k): \tau_{j,k}=\tau^*\},
\end{align} where the maximum is in lexicographic order. %Note from~\eqref{eq37a} that we always can choose a ``smallest'' pair $(j,k)$ since there exists at least one pair $(j,k)$ is such that $\tau_{j,k}=\tau^*$.
%\end{itemize}

Let $X_1^{\infty},X_2^{\infty},\bar{X}_1^{\infty},\bar{X}_2^{\infty},Y^{\infty}$ be i.i.d.\ infinite-dimensional vectors with joint distribution
\begin{align}
&\bbP_{X_1  X_2  Y \bar{X}_1 \bar{X}_2 }(x_1,x_2,y,\bar{x}_1,\bar{x}_2)\nn\\
&=\bbP_{X_1}(x_1) \bbP_{X_2}(x_2) \bbP(y|x_1 x_2) \bbP_{X_1}(\bar{x}_1)\bbP_{X_2}(\bar{x}_2),
\end{align}
 where $\bbP_{X_1} \sim \calN(0,P_1)$, $\bbP_{X_2} \sim \calN(0,P_2)$ and $\bbP(y|x_1  x_2)$ is the law of the Gaussian MAC.  

For each finite $n$, define three  random  information density random variables (random walks) $S_n^{(1)}:=i(X_1^n;Y^n|X_2^n)$, $S_n^{(2)}:=i(X_2^n;Y^n|X_1^n)$, and $S_n^{(3)}:=i(X_1^n,X_2^n;Y^n)$   and hitting times
\begin{align}
\tau^{(1)}&:=\inf\{n\geq 0: i(X_1^n;Y^n|X_2^n) > \gamma_1\},\label{eqn:tau_1}\\
\tau^{(1)}_+&:=\inf\{n\geq 0: i(X_1^n;Y^n|X_2^n) >0\},\label{eqn:tau_plus1}\\
\bar{\tau}^{(1)}&:=\inf\{n\geq 0: i(\bar{X}_1^n;Y^n|X_2^n) > \gamma_1\},
\end{align}
Analogously define $\tau^{(2)}$, $\tau^{(2)}_+$, $\bar{\tau}^{(2)}$, $\tau^{(3)}$, $\tau^{(3)}_+$, $\bar{\tau}^{(3)}$ and
$
 \tau' :=\max\{\tau^{(1)},\tau^{(2)},\tau^{(3)}\}.
$
%\\
%\tau^{(2)}&:=\inf\{n\geq 0: i(X_2^n;Y^n|X_1^n) > \gamma_2\},\label{eqn:tau_2}\\
%\tau^{(2)}_+&:=\inf\{n\geq 0: i(X_2^n;Y^n|X_1^n) > 0\},\\
%\bar{\tau}^{(2)}&:=\inf\{n\geq 0: i(\bar{X}_2^n;Y^n|X_1^n) > \gamma_2\},\\
%\tau^{(3)}&:=\inf\{n\geq 0: i(X_1^n,X_2^n;Y^n) > \gamma_3\},\label{eqn:tau_3}\\
%\tau^{(3)}_+&:=\inf\{n\geq 0: i(X_1^n,X_2^n;Y^n) > 0\},\\
%\bar{\tau}^{(3)}&:=\inf\{n\geq 0: i(\bar{X}_1^n,\bar{X}_2^n;Y^n) > \gamma_3\},\\
%\label{eq60b}
%\textcolor{red}{\tau'}&:=\max\{\tau^{(1)},\tau^{(2)},\tau^{(3)}\}.
%\end{align} 

It follows that the average length of the transmission satisfies
\begin{align}
\bbE(\tau^*)&=\frac{1}{M_1M_2}\sum_{j,k} \bbE(\tau^*|W_1=j,W_2=k)\\*
&= \bbE(\tau^*|W_1=1,W_2=1)\\
&\leq \bbE(\max\{\tau^{(1)}_{1,1},\tau^{(2)}_{1,1},\tau^{(3)}_{1,1}\}|W_1=1,W_2=1)\\
\label{eq55a}
&=\bbE(\max\{\tau^{(1)},\tau^{(2)},\tau^{(3)}\})=\bbE(\tau').
\end{align}
From the analysis of the DM-MAC in Trillingsgaard and Popovski~\cite{Trillingsgaard:2014}, we know that the average probability of error satisfies
\begin{align}
\bbP(g(Y^{\tau^*}) \neq (W_1,W_2))&\leq (M_1-1)(M_2-1) \bbP(\tau' \geq \bar{\tau}^{(3)})\nn\\*
&\quad +(M_1-1) \bbP(\tau' \geq \bar{\tau}^{(1)})\nn\\*
&\quad +(M_2-1) \bbP(\tau' \geq \bar{\tau}^{(2)}). \label{eqn:error_pr}
\end{align}
%We define random walks $S_n^{(1)} :=i(X_1^n;Y^n|X_2^n), S_n^{(2)} :=i(X_2^n;Y^n|X_1^n)$, and $S_n^{(3)} :=i(X_1^n,X_2^n;Y^n)$. 
Observe that the following statistics are all finite:
\begin{align}
\mu_1&=\bbE[i(X_1;Y|X_2)]=I(X_1;Y|X_2)=\rvC(P_1),\\
%\mu_2&=\bbE[i(X_2;Y|X_1)]=I(X_2;Y|X_1)=\rvC(P_2),\\
%\mu_3&=\bbE[i(X_1,X_2;Y)]=I(X_1,X_2;Y)=\rvC(P_1+P_2),\\
\sigma_1^2&=\var\left(i(X_1;Y|X_2)\right)=\frac{P_1}{1+P_1},
\end{align} \vspace{-.25in}
%\sigma_2^2&=\var\left(i(X_2;Y|X_1)\right)=\frac{P_2}{1+P_2},\\
%\sigma_3^2&=\var\left(i(X_1,X_2;Y)\right)=\frac{P_1+P_2}{1+P_1+P_2},\\
\begin{align}
 \! \! \! \!\bbE\left[i(X_1;Y|X_2)^+\right]&\leq \bbE\left[i(X_1;Y|X_2)^+ \! + \! i(X_1;Y|X_2)^- \right]\\
&=\bbE\left[|i(X_1;Y|X_2)|\right]\\
&\le \sqrt{\bbE\left[(i(X_1;Y|X_2))^2\right]}\\
&\le \sqrt{\mu_1^2+\sigma_1^2} <\infty.
%\bbE\left[i(X_2;Y|X_1)^+\right]&\leq \sqrt{\mu_2^2+\sigma_2^2} <\infty,\\
%\bbE\left[i(X_1,X_2;Y)^+\right]&\leq \sqrt{\mu_3^2+\sigma_3^2} <\infty.
\end{align} 
Similarly, $\mu_2=\bbE[i(X_2;Y|X_1)]=I(X_2;Y|X_1)$, $\mu_3=\bbE[i(X_1,X_2;Y)]=I(X_1,X_2;Y)$, $\sigma_2^2=\var\left(i(X_2;Y|X_1)\right)$, $\sigma_3^2=\var\left(i(X_1,X_2;Y)\right)$, $\bbE\left[i(X_2;Y|X_1)^+\right]$, and $\bbE\left[i(X_1,X_2;Y)^+\right]$ are finite. 
Moreover, by~\cite[pp.~207]{Battacharya}, Cramer's condition (C) in Definition~\ref{stronglattice} is satisfied by those distributions having at least a continuous component in its Lebesgue decomposition. Since $i(X_1,Y|X_2)$,   $i(X_2;Y|X_1)$, and $i(X_1,X_2;Y)$ are all continuous random variables, their distribution functions are strongly nonlattice. Hence, it follows from~Lemma~\ref{lem00} that
\begin{align}
\label{eq309newest}
I(X_1;Y|X_2) \bbE(\tau^{(1)})&=\gamma_1+ \xi_1+o(1), \; \mbox{as} \; \gamma_1 \to \infty,\\
I(X_2;Y|X_1) \bbE(\tau^{(2)})&=\gamma_2+ \xi_2+o(1), \; \mbox{as} \; \gamma_2 \to \infty,\\
\label{eq311newest}
I(X_1,X_2;Y) \bbE(\tau^{(3)})&=\gamma_{3}+\xi_3+o(1), \; \mbox{as} \; \gamma_3 \to \infty,
\end{align}
where
\begin{align}
\xi_j :=  \frac{\bbE  \Big[ \Big(S^{(j)}_{\tau_+^{(j)}} \Big)^2 \Big] }{2\bbE \Big[ S^{(j)}_{\tau_+^{(j)}} \Big]},\quad j=1,2,3.
\end{align}
Recall that $S_n^{(1)}$ is the $n$-letter information density $i(X_1^n;Y^n|X_2^n)$ and $\tau_+^{(1)}$ is defined in \eqref{eqn:tau_plus1}. Additionally,  let
\begin{align}
\nu_j  &:=  \frac{\bbE  \Big[ \Big(S^{(j)}_{\tau_+^{(j)}} \Big)^3 \Big] }{  \bbE \Big[ S^{(j)}_{\tau_+^{(j)}} \Big]},\quad\mbox{and}\\
\tau^{(j)}(x)&:=\inf\{n\geq 0: S_n^{(j)} >x\},\quad j=1,2,3. 
\end{align}
%In addition, from Lemma~\ref{lem0} we also have for $j=1,2,3$ that
%Recall the definitions of  $\tau^{(j)}, j = 1,2,3$ in \eqref{eqn:tau_1}, \eqref{eqn:tau_2} and \eqref{eqn:tau_3}. 
From Lemma~\ref{lem0},  we have  that
\begin{align}
\var(\tau^{(j)})&=\mu_j^{-3} \sigma_j^2 \gamma_j +\mu_j^{-2}K_j +o(1),\; \mbox{as}\;\gamma_j \to \infty, \label{eqn:vari}
\end{align}
where for $ j=1,2,3$, %$\tau^{(j)}, j = 1,2,3$ are defined in \eqref{eqn:tau_1}, \eqref{eqn:tau_2} and \eqref{eqn:tau_3} and
%\begin{align}
%K_j&= \frac{\sigma_j^2 \bbE S_{\tau_+^{(j)}}^2}{2\mu_1 \bbE S_{\tau_+^{(j)}}}+\frac{3}{4}\left(\frac{\bbE S_{\tau_+^{(j)}}^2}{\bbE S_{\tau_+^{(j)}}}\right)^2 -\frac{2}{3} \left(\frac{\bbE S_{\tau_+^{(j)}}^3}{\bbE S_{\tau_+^{(j)}}}\right)
% - \left(\frac{\bbE S_{\tau_+^{(j)}}^2}{\bbE S_{\tau_+^{(j)}}}\right) \bbE\{\min_{n\geq 0}S^{(j)}_n\}\nn\\
%&\quad -2\int_0^{\infty} \bbE\{S^{(j)}_{\tau^{(j)}(x)}-x\}\bbP\{\min_{n\geq 0} S^{(j)}_n \leq -x\} \, \rmd x, \quad  j=1,2,3
%\end{align}
%\begin{align}
%K_j&:=\frac{\sigma_j^2}{\mu_j} \frac{\bbE  \Big[ \Big(S^{(j)}_{\tau_+^{(j)}} \Big)^2 \Big] }{2\bbE \Big[ S^{(j)}_{\tau_+^{(j)}} \Big]} + 3 \left( \frac{\bbE  \Big[ \Big(S^{(j)}_{\tau_+^{(j)}} \Big)^2 \Big] }{2\bbE \Big[ S^{(j)}_{\tau_+^{(j)}} \Big]}\right)^2 - \frac{2}{3} \left(  \frac{\bbE  \Big[ \Big(S^{(j)}_{\tau_+^{(j)}} \Big)^3 \Big] }{2\bbE \Big[ S^{(j)}_{\tau_+^{(j)}} \Big]} \right) - \left( \frac{\bbE  \Big[ \Big(S^{(j)}_{\tau_+^{(j)}} \Big)^2 \Big] }{2\bbE \Big[ S^{(j)}_{\tau_+^{(j)}} \Big]}\right) \bbE\{\min_{n\geq 0}S^{(j)}_n\}\nn\\
% &\qquad -2\int_0^{\infty} \bbE\{S^{(j)}_{\tau^{(j)}(x)}-x\}\bbP\{\min_{n\geq 0} S^{(j)}_n \leq -x\} \, \rmd x, \quad  j=1,2,3
%\end{align}
\begin{align}
K_j &:= \frac{\sigma_j^2}{\mu_j} \xi_j + 3\xi_j^2 -\frac{2 }{3}\nu_j-2\xi_j\, \bbE\Big\{\min_{n\geq 0}S^{(j)}_n\Big\}\nn\\
 &\; -2\int_0^{\infty} \bbE\{S^{(j)}_{\tau^{(j)}(x)}-x\}\bbP\Big\{\min_{n\geq 0} S^{(j)}_n \leq -x\Big\} \, \rmd x.
\end{align}
 are constants which are not dependent on $\gamma_j,j=1,2,3,$ (i.e. $K_1,K_2,K_3=O(1)$).
%and
%\begin{align}
%\tau^{(j)}(x)&=\inf\{n\geq 0: S_n^{(j)} >x\},\quad j=1,2,3.
%\end{align}
Now, for any positive real number $N'$, choose
\begin{align}
\label{eq102b}
\gamma_1&=I(X_1;Y|X_2)(N'-A\sqrt{N'}-\cD),\\
\label{eq104}
\gamma_2&=I(X_2;Y|X_1)(N'-A\sqrt{N'}-\cD),\\
\label{eq105}
\gamma_3&=I(X_1,X_2;Y)(N'-A\sqrt{N'}-\cD),
\end{align}
and a pair $(M_1,M_2)$ satisfying 
\begin{align}
0 \leq \log M_j &\leq \gamma_j-\log(3N'), \quad j=1,2, \label{eqn:choiceMj}\\
0\le\log M_1 M_2 &\leq \gamma_3-\log(3N'),\label{eqn:choiceMjj}
\end{align} for some $A\geq 0, \cD\geq 0$ to be determined later. These choices of $M_1$ and $M_2$ and the fact that $\xi_j=O(1)$ for  all $j=1,2,3$ show that  \eqref{eq213newq} and \eqref{eq215newq} are satisfied.

Combining these choices of $\gamma_j$ with~\eqref{eq309newest}--\eqref{eq311newest} we obtain 
\begin{align}
\label{eq101b}
\bbE[\tau^{(j)}]&= N'-A\sqrt{N'}-\cD-B_j+o(1) \quad j=1,2,3,
\end{align}
where
\begin{align}
B_j&:=\frac{-2\xi_j}{\log\left(1+ P_j\right)} , \quad j=1,2\\
B_3&:=\frac{-2\xi_3}{\log\left(1+ P_1+P_2\right)} ,
\end{align} are constants. By using the facts that $A\ge 0$, $\cD\ge 0$ and \eqref{eqn:vari}, we also have
\begin{align}
\var(\tau^{(j)})&= \cR_j (N'-A\sqrt{N'}-\cD)+\cS_j+o(1)  \nn\\*
&\leq \cR_j N'+\cS_j+o(1),
\end{align} where the constants $\cR_j$ and $\cS_j$ are defined according to Lemma~\ref{lem0}. Specifically, 
\begin{align}
\cR_j&:=\left(\frac{\sigma_j}{\mu_j}\right)^2=\eqref{para_rnewnew}, \\%\quad j=1,2,3\\
\cS_j&:=\mu_j^{-2} K_j, \quad j=1,2,3.
\end{align}
It follows from Lemma~\ref{lem1b} that
\begin{align}
\label{eqnew2017}
\bbE(\tau^*) \leq \bbE[\tau']=\bbE[\max\{\tau^{(1)},\tau^{(2)},\tau^{(3)}\}] \le N'+o(1)
\end{align}
 as  $N' \to \infty$. 
%if we choose $A$ as~\eqref{para_anewnew}, and $\cD$ sufficiently large.
Moreover, from~\eqref{eq101b} we have $\bbE(\tau^{(j)})<\infty$, hence
\begin{align}
\bbP(\tau^{(j)}<\infty)&=1,\quad j=1,2,3. \label{eqn:Ptauj}
\end{align}
Applying a change of measure, we observe that for any measurable function $f$,
\begin{align}
\label{eq110c}
\bbE[f(\bar{X}_1^n,X_2^n,Y^n)]& \!=\!\bbE\big[f(X_1^n,X_2^n,Y^n) \exp\big(\!-\! S_n^{(1)}\big)\big],\\
\bbE[f(X_1^n,\bar{X}_2^n,Y^n)]&\!=\!\bbE\big[f(X_1^n,X_2^n,Y^n) \exp\big(\!-\! S_n^{(2)}\big)\big],\\
\label{eq351lan}
\bbE[f(\bar{X}_1^n,\bar{X}_2^n,Y^n)]&\!=\!\bbE\big[f(X_1^n,X_2^n,Y^n) \exp\big(\!-\! S_n^{(3)}\big)\big].
\end{align}
Observe that
$
1\{\tau^{(j)}\leq n\} \in \sigma(X_1^n,X_2^n,Y^n)$ for $j=1,2,3$, $
1\{\tau' \leq n\}\in \sigma(X_1^n,X_2^n,Y^n)$,
$1\{\bar{\tau}^{(1)} \leq n\} \in \sigma(\bar{X}_1^n,X_2^n,Y^n)$,
$1\{\bar{\tau}^{(2)} \leq n\}  \in \sigma(X_1^n,\bar{X}_2^n,Y^n)$,  and
$1\{\bar{\tau}^{(3)} \leq n\} \in \sigma(\bar{X}_1^n,\bar{X}_2^n,Y^n)$.
Following the same arguments as in~\cite[Eqns.~(111)--(118)]{Yury2011}, we have
\begin{align}
\bbP(\bar{\tau}^{(3)}\leq \tau')&\leq \bbP(\bar{\tau}^{(3)}<\infty)
%&\leq \bbP(\bar{\tau}^{(3)}<\infty)\\
%&=\lim_{n\to \infty} \bbP(\bar{\tau}^{(3)}<n)\\
%%&\stackrel{(a)}{=}\lim_{n\to \infty} \bbP\left(\exp\left(-S_n^{(3)}\right)1\{\tau<n\}\right)\\
%&\stackrel{(a)}{\leq}\lim_{n\to \infty} \bbE\left(\exp\left(-S_n^{(3)}\right)1\{\tau^{(3)}<n\}\right)\\
%&\stackrel{(b)}{=} 
\le\exp(-\gamma_3),\label{eqn:bound_tau0} 
%\bbP(\bar{\tau}^{(1)}\leq \tau)&\leq \bbP(\bar{\tau}^{(1)}<\infty)\\
%&\leq \exp(-\gamma_1).
\end{align} %Here, (a) follows from the fact that $1\{\bar{\tau}^{(3)} \leq n\} \in \sigma(\bar{X}_1^n,\bar{X}_2^n,Y^n), 1\{\tau^{(3)}\leq n\} \in \sigma(X_1^n,X_2^n,Y^n)$ and the change of measure formula~\eqref{eq351lan}, (b) \textcolor{red}{follows from the definition of $\tau^{(3)}$ and~\cite[Equation~(118)]{Yury2011}}.  
Similarly, for $j=1,2$,
\begin{align}
\bbP(\bar{\tau}^{(j)}\leq \tau')&\leq \bbP(\bar{\tau}^{(j)}<\infty)\leq \exp(-\gamma_j). \label{eqn:bound_tau1}
\end{align}
From  the bound on the error probability in~\eqref{eqn:error_pr}, the bounds on the individual probabilities in~\eqref{eqn:bound_tau0} and \eqref{eqn:bound_tau1}, the choices of $M_1$ and $M_2$ in~\eqref{eqn:choiceMj}  and \eqref{eqn:choiceMjj}, we see that the average error probability of the stop-feedback code satisfies 
\begin{align}
\eps'  \le\frac{1}{N'}\label{expnt}. 
% %&\leq  (M_1-1)(M_2-1) \bbP(\bar{\tau}^{(3)}\leq \tau) \nn\\*
%%&\qquad + (M_2-1) \bbP(\bar{\tau}^{(2)}\leq \tau) \nn\\*
%%&\qquad  + (M_1-1) \bbP(\bar{\tau}^{(1)}\leq \tau)\\
%&\leq (M_1-1)(M_2-1) \exp(-\gamma_{3})\nn\\*
%&\quad + (M_1-1) \exp(-\gamma_{1})\nn\\*
%&\quad+ (M_2-1)\exp(-\gamma_{2})\leq \frac{1}{N'}.
\end{align}
Observe that 
\begin{align}
\bbE\bigg[\sum_{n=1}^{\tau^*}X_{jn}^2 \bigg]&=\bbE \bigg[\sum_{n=1}^{\tau^*}X_{jn}^2 \,\bigg|\,W_1=1,W_2=1\bigg]\\
&\leq \bbE\bigg[\sum_{n=1}^{\tau_{1,1}}X_{jn}^2\,\bigg|\,W_1=1,W_2=1\bigg]\\
%&=\bbE\bigg[\sum_{n=1}^{\max\{\tau_{1,1}^{(1)}, \tau_{1,1}^{(2)},\tau_{1,1}^{(3)}\}}X_{jn}^2|W_1=1,W_2=\bigg]\\
\label{power:bound2}
&=\bbE\bigg[\sum_{n=1}^{\tau'}X_{jn}^2\bigg],\quad j=1,2.
\end{align}
To verify that the expected power constraints are satisfied,  we now check all the conditions of Lemma~\ref{lem3a}  (with $X_{jn}^2$ for $j=1,2$ here playing the role of $X_n$ in Lemma \ref{lem3a}).
\begin{itemize}
\item We have  
$
\bbE[X_{jn}^2]=P_j$ for $j=1,2$
 so it follows that $X_{1n}^2$ and $X_{2n}^2$ are integrable for all $n\ge 1$.
\item Now, we see that $1\{\tau' \geq n\}=1-1\{\tau' \leq n-1\} \in \sigma(X_1^{n-1},X_2^{n-1},Y^{n-1})$. Moreover, since the sequence $\{X_{1n}\}_{n\geq 1}$ as well as the sequence $\{X_{2n}\}_{n\geq 1}$ are i.i.d.\ generated and the channel is memoryless, we have  that $1\{\tau' \geq n\}$ is independent of $X_{1n}$ and $X_{2n}$. It follows that
$\bbE[X^2_{jn} 1\{\tau' \geq n\}]=\bbE[X^2_{jn}]\bbE[1\{\tau' \geq n\}]
=\bbE[X^2_{jn}]\bbP(\tau' \geq n)$ for $j=1,2$;
\item For each $j=1,2$, the infinite series $\sum_{n=1}^{\infty} \bbE[X_{jn}^21\{\tau' \geq n\}]$ satisfies
\begin{align}
\sum_{n=1}^{\infty} \bbE[X_{jn}^21\{\tau' \geq n\}]&=\sum_{n=1}^{\infty} \bbE[X_{jn}^2]\bbP(\tau' \geq n)\\
&\leq P_j \sum_{n=1}^{\infty} \bbP(\tau' \geq n)\\
&=P_j \bbE(\tau' )\\
\label{eqn:new1}
&\leq P_j (N'+o(1)) <\infty,
\end{align} where~\eqref{eqn:new1} follows from~\eqref{eqnew2017}.
\item For each $j=1,2$, all random variables $X^2_{jn}, n\ge 1$ have the same expectation $P_j$.
\item $\bbE(\tau') \leq N'+o(1) <\infty$.
\end{itemize}
Hence, by~\eqref{power:bound2} and Lemma~\ref{lem3a}, the expected power constraints at the encoders satisfy
\begin{align}
\bbE\bigg[\sum_{n=1}^{\tau^*}X_{jn}^2\bigg] &\leq \bbE\bigg[\sum_{n=1}^{\tau'} X_{jn}^2\bigg]\\
&=\bbE(\tau') \bbE[X^2_{j1}]\\
&\le\bbE(\tau' ) P_j, \quad j=1,2. \label{expec:new}
\end{align} 
This means that we have shown there exists an $(M_1,M_2,N'+o(1),\frac{1}{N'})$ stop-feedback code with stopping time $\tau^*$ such that~\eqref{expec:new} holds. Since there exists such a  code, we can find an $(M_1,M_2,N'+o(1),\frac{1}{N'})$ stop-feedback code with stopping time $\tau'$ by increasing the stopping time from $\tau^*$ to $\tau'$ (using the same decoder at time $\tau^*$). It follows that~\eqref{expec:new} holds with equality.  Moreover, if there exists an $(M_1,M_2,N'+o(1),\frac{1}{N'})$ stop-feedback code with stopping time $\tau^*$, by keeping the same stopping rule and  the decoder  of the aforementioned  code and setting 
\begin{align}
\tilX_{jn} &:=\begin{cases}X_{jn},&n\leq \tau^*\\0,&n> \tau^*\end{cases}, \quad j=1,2,
\end{align}% where ${\bf C}^{(j)}_{w_j,n}$ is the $n$-th coordinate of the vector ${\bf C}^{(j)}_{w_j}$. Then 
we have a new $(M_1,M_2,N'+o(1),\frac{1}{N'})$ stop-feedback code satisfying:
\begin{align}
\label{eqn:new2}
\sum_{n=1}^{\infty} \bbE[\tilX_{jn}^2] &=\bbE\bigg[\sum_{n=1}^{\infty} \tilX_{jn}^2\bigg]\\
&=\bbE\bigg[\sum_{n=1}^{\tau^*} X_{jn}^2\bigg]\\
&=\bbE(\tau^*) P_j, \quad j=1,2,
\end{align} 
%The calculation here is analogous to \eqref{eqn:sf_calc1}--\eqref{eqn:sf_calc3}.
where~\eqref{eqn:new2} follows from Tonelli's theorem~\cite{Billingsley}. This concludes the proof of Lemma \ref{lem1}.
\end{IEEEproof}
\begin{lemma}
\label{them6}
For the Gaussian MAC $\bbP(y|x_1,x_2)$, there exists an $(M_1,M_2,N,P_1,P_2,\eps)$ stop-feedback code for $M_1,M_2$ satisfying~\eqref{eq166newq} and \eqref{eq167newq}.% and
%\begin{align}
%0  \leq \log M_j &\leq \left(\frac{N}{1-\eps}-A\sqrt{\frac{N}{1-\eps}}\right)\rvC(P_j)-\log N+O(1),\quad j=1,2 \label{eqn:boundMj}\\*
%0 \leq \log M_1M_2& \leq \left(\frac{N}{1-\eps}-A\sqrt{\frac{N}{1-\eps}}\right)\rvC(P_1+P_2)-\log N+O(1).\label{eqn:boundM1M2}
%\end{align} Here, 
%$A$ is defined in $\eqref{para_anewnew}$.
\end{lemma}
\begin{IEEEproof}
We propose a stop-feedback coding scheme as follows:
\begin{itemize}
\item The decoder chooses  numbers $N',P_1',P_2'$ such that
\begin{align}
\label{eq62n}
\frac{(N')^2(1-\eps)}{N'-1}&\leq N,\\
\label{eq63n}
P_j'&= P_j,\quad j=1,2.
\end{align}
\item The decoder generates a Bernoulli random variable $D\sim \mathrm{Bern}(p)$, where 
\begin{align}
p:=\frac{N'\eps-1}{N'-1}.\label{eqn:Bernp}
\end{align}
\item If $D=1$, the decoder sends a stop-feedback (or a NACK) to the encoder via the feedback link. This means that $\tau=0$.
\item If $D=0$, the encoder  sends the intended message to the decoder using the stop-feedback $(M_1,M_2,N'+o(1),P'_1,P_2',\frac{1}{N'})$ mentioned in   Lemma~\ref{lem1} for the Gaussian MAC with expected powers $P'_1$ and $P'_2$ and stops at time $\tau'$. This means that $\tau=\tau'$. 
\end{itemize}
It follows that the error probability of the proposed stop-feedback coding scheme is upper bounded by
\begin{align}
1\frac{N'\eps-1}{N'-1}+\left(1-\frac{N'\eps-1}{N'-1}\right)\frac{1}{N'}=\eps. \label{eqn:err_prob_sf_mac}
\end{align}
In addition, the average length of the proposed   stop-feedback coding scheme is less than or equal to
\begin{align}
& \left(1-\frac{N'\eps-1}{N'-1}\right)\bbE(\tau')\nn\\*
 &\leq  \left(1-\frac{N'\eps-1}{N'-1}\right) N'+o(1) \left(1-\frac{N'\eps-1}{N'-1}\right)\\
&=\frac{(N')^2(1-\eps)}{N'-1}+o(1)\\
&\leq N+o(1).\label{eqn:length_sf_mac}
\end{align}
From~\eqref{eq62n} and~\eqref{eq63n}, the expected powers of the combined scheme satisfy
\begin{align}
\left(1-\frac{N'\eps-1}{N'-1}\right)\bbE(\tau') P'_j = \bbE(\tau) P'_j=\bbE(\tau) P_j,\; j=1,2. \label{eqn:powers2_sf_mac}
\end{align}
Therefore, combining this code construction with Lemma~\ref{lem1}, we see that there exists an  $(M_1,M_2,N+o(1),P_1,P_2,\eps)$ stop-feedback code where
\begin{align}
0\leq \log M_j &\leq \left(\frac{N}{1-\eps}-A\sqrt{\frac{N}{1-\eps}}-\cD+o(1)\right)\rvC(P_j) \nn\\*
&\quad-\log\left(\frac{N}{1-\eps}\right)+O(1),\quad j=1,2\\
0\le \log M_1M_2 &\leq \left(\frac{N}{1 \! - \! \eps} \! - \! A\sqrt{\frac{N}{1 \! - \! \eps}} \! - \! \cD \! + \! o(1)\right)\rvC(P_1 \! + \! P_2) \nn\\*
&\quad-\log\left(\frac{N}{1-\eps}\right)+O(1).
\end{align}
Observe that if there exists an $(M_1,M_2,N+o(1),P_1,P_2,\eps)$ stop-feedback code, then there also exists an $(M_1,M_2,N,P_1,P_2,\eps)$ stop-feedback code by setting the expected length equal to $N-o(1)$. This change of the expected length does not affect the asymptotic approximation of the code rates. This concludes our proof of the achievability part of Theorem~\ref{stop_mac}.\end{IEEEproof}
\subsection{Achievability Proof for Theorem~\ref{vlft_mac}}\label{sec:ach_v}
\begin{lemma} \label{lem:vlft}
Given a Gaussian MAC, for any $\rho \in [0,1]$, there exist an $(M_1,M_2,N,P_1,P_2,\eps)$ VLFT-feedback code for any $M_1, M_2$ satisfying \eqref{eqn:vlft_low1} and \eqref{eqn:vlft_low3}.
%\begin{align}
%0\leq \log M_j &\leq \left(\frac{N}{1-\eps}\right)\rvC\left(P_j(1-\rho^2)\right)-\log \log N +O(1),\quad j=1,2\\*
%0\le\log M_1 M_2 &\leq \left(\frac{N}{1-\eps}\right)\rvC\left(P_1+P_2+2\rho \sqrt{P_1P_2} \right)-\log \log N+O(1).
%\end{align}
\end{lemma}
\begin{IEEEproof}
Consider Ozarow's coding scheme (for the Gaussian MAC with fixed-length feedback)~\cite{Ozarow} with fixed blocklength $N'\in\bbN$, expected powers bounded by $P'_1$ and $P'_2$, and message sizes $M_1$ and $M_2$ satisfying
\begin{align}
\label{eq560new}
\log M_j&=N' \rvC\left(P'_j(1-\rho^2)\right) \nn\\*
&\qquad-\log \log N' +O(1), \quad j=1,2\\
 \log M_1 M_2&=N' \rvC\big(P_1+P_2+2\rho\sqrt{P_1P_2}\big)\nn\\*
 &\qquad-\log \log N' +O(1) \label{eq:eq162rev}
\end{align} where $\rho \in [0,1]$.  Then,  from~\cite[Eqn.~(13)]{Ozarow} and \cite[Eqn.~(121)]{TruongFongTan17}, one sees that  Ozarow's scheme results in an  error probability
\begin{align}
\eps' &\leq \frac{2}{(N')^2}\leq \frac{1}{N'},\quad \forall\, N'\ge 2.
\end{align} %for $N'\geq 2$ (cf.~\cite[Equation~(121)]{TruongFongTan17}).

Therefore, we   construct the VLFT coding scheme as follows.
\begin{itemize}
\item The decoder chooses the largest natural number $N'$ such that \eqref{eq62n} is satisfied. It also chooses positive numbers $P'_1,P_2'$ as in~\eqref{eq63n}.
\item The decoder generates a Bernoulli random variable $D\sim \mathrm{Bern}(p)$, where $p$ is defined in~\eqref{eqn:Bernp}.
\item If $D=1$, the decoder sends a stop-feedback signal (or a NACK) to the encoder via the feedback link. This means that, conditioned on $D=1$, $\tau=0$.
\item If $D=0$, the encoder  sends the intended message to the decoder using   Ozarow's coding scheme with parameters $(M_1,M_2,N',P'_1,P_2',\frac{1}{N'})$  with expected powers $P'_1=P_1$ and $P'_2=P_2$ and stops at time $\tau'$. This means that, conditioned on $D=0$, we have $\tau=N'$. 
\end{itemize}
Similarly to the stop-feedback case, it follows that the error probability of the proposed VLFT   coding scheme is upper bounded by $\eps$.  The expected powers of the combined scheme are also bounded by $\bbE(\tau)P_j,j=1,2$. Consequently, the achievability part of Theorem~\ref{vlft_mac} is proved. %follows from~\eqref{eq560new},~\eqref{eq62n}, and~\eqref{eq63n}.
\end{IEEEproof}
\section{Converse Proofs}\label{sec:conv}
\subsection{Converse Proof   for Theorem~\ref{stop_mac}}\label{sec:conv_s}
\begin{lemma} \label{lem:convThm3}
Given a Gaussian MAC $\bbP(y|x_1,x_2)$, $0\leq \eps\leq 1-\max\{\frac{1}{M_1},\frac{1}{M_2}\}$, any $(M_1,M_2,N,P_1,P_2,\eps)$ stop-feedback code satisfies \eqref{eqn:sf_up1} and \eqref{eqn:sf_up3}   for all $N\in\bbN$.
%\begin{align}
%\label{conv1_stop}
%0\leq \log M_j &\leq \frac{N\rvC(P_j)+h_{\rmb}(\eps)}{1-\eps},\quad j=1,2\\
%\label{conv3_stop}
%0\le\log M_1 M_2 & \leq \frac{N\rvC(P_1+P_2)+h_{\rmb}(\eps)}{1-\eps}.
%\end{align}
\end{lemma}
\begin{IEEEproof}
First, we consider the case $|\calU_1|=|\calU_2|=1$. For the stop-feedback formalism, $\tau$ is a stopping time of the filtration $\{\sigma(Y^n)\}_{n=0}^{\infty}$. We note that if there exists a code $(f_n^{(1)},f_n^{(2)},g_n,\tau)$, we can construct another code $(\hat{f}_n^{(1)},\hat{f}_n^{(2)},\hat{g}_n,\hat{\tau})$ such that $\hat{X}_n=\hat{Y}_n=\rvT$ for any $n\geq \hat{\tau}$, where $\rvT \notin \bbR$ is a special symbol appended to the input and output alphabets to form the common input-output alphabet $\bbR\cup\{\rvT\}$ and $\hat{\tau}=\tau+1=\inf\{n: \hat{Y}_n=\rvT\}$. Thus for the converse, it is suffices to consider $(\hat{f}_n^{(1)},\hat{f}_n^{(2)},\hat{g}_n,\hat{\tau})$, where the encoders $\hatf_n^{(j)}, j = 1,2$ are defined as in~\cite[Eqn.~(59)]{Yury2011} and the decoder $\hat{g}_n$ as in~\cite[Eqn.~(61)]{Yury2011}. 

%Define
%\begin{align}
%\label{eq129a}
%\hat{g}_n(\hat{Y}^n):=\begin{cases} g_n(\hat{Y}^n),&n < \hat{\tau}\\g_n(Y^{\hat{\tau}-1}),&n \geq \hat{\tau} \end{cases}.
%\end{align}
 In addition, using the same arguments as~\cite[Eqn.~(68)]{Yury2011} we have
\begin{align}
\label{eq418newl}
(1-\eps)\log M_1M_2 &\leq I(W_1W_2;\hat{Y}^{\infty})+h_{\rmb}(\eps),\\
\label{eq419newl}
(1-\eps)\log M_1 &\leq I(W_1;\hat{Y}^{\infty}|W_2=w_2)+h_{\rmb}(\eps),\\
\label{eq420newl}
(1-\eps)\log M_2 &\leq I(W_2;\hat{Y}^{\infty}|W_1=w_1)+h_{\rmb}(\eps).
\end{align}
By taking expectations of~\eqref{eq419newl} and~\eqref{eq420newl} with respect to $P_{W_2}$ and $P_{W_1}$ respectively, we obtain
\begin{align}
\label{eq171n}
(1-\eps)\log M_1 &\leq I(W_1;\hat{Y}^{\infty}|W_2)+h_{\rmb}(\eps),\\
\label{eq173n}
(1-\eps)\log M_2 &\leq I(W_2;\hat{Y}^{\infty}|W_1)+h_{\rmb}(\eps).
\end{align}
Define 
\begin{align}
\Psi_n:=1\{\hat{\tau} \leq n-1\} \in \sigma(\hat{Y}^{n-1}). \label{eqn:Psi}
\end{align}
By   Lemma~\ref{lem3} in   Appendix \ref{app:mac}, we have
\begin{align}
\label{eq127new}
&I(W_1W_2;\hat{Y}^{\infty})\nn\\*
&\quad\leq \sum_{n=1}^{\infty} \frac{1}{2}\bbP(\Psi_n=0)\log(1+\bbE[(X_{1n}+X_{2n})^2|\Psi_n=0]) ,\\
\label{eq128new}
& I(W_1;\hat{Y}^{\infty}|W_2)\nn\\* 
&\quad\leq \sum_{n=1}^{\infty} \frac{1}{2} \bbP(\Psi_n=0)\log(1+\bbE[X_{1n}^2|\Psi_n=0]),\\
\label{eq129new}
&I(W_2;\hat{Y}^{\infty}|W_1)\nn\\*
&\quad\leq \sum_{n=1}^{\infty} \frac{1}{2} \bbP(\Psi_n=0)\log(1+\bbE[X_{2n}^2|\Psi_n=0]).
\end{align}
We observe that
\begin{align}
\sum_{n=1}^{\infty} \bbP(\Psi_n=0)=\sum_{n=1}^{\infty} \bbP(\tau\geq n)
%&=\sum_{n=1}^{\infty} \bbP(\tau\geq n)\\
=\bbE(\tau).
\end{align}
It follows that
%\begin{align}
%\sum_{n=1}^{\infty} \frac{\bbP(\Psi_n=0)}{\bbE(\tau)}=1, \label{eqn:prob_dist}
%\end{align}
%so 
$\{ {\bbP(\Psi_n=0)}/{\bbE(\tau)}\}_{n=1}^\infty$ is a probability distribution. 
Moreover, since the function $f(x)=\log(1+x)$ is concave, we have from~\eqref{eq418newl} and~\eqref{eq127new} that
\begin{align}
&(1-\eps)\log M_1 M_2 \nn\\*
 &\leq \frac{1}{2} \bbE(\tau)  \log\left(1+\sum_{n=1}^{\infty} \frac{\bbP(\Psi_n=0)}{\bbE(\tau)} \bbE[(X_{1n}\! +\! X_{2n})^2|\Psi_n\! =\! 0]\right)\nn\\*
 &\qquad+h_{\rmb}(\eps) \label{eqn:setineq1}\\
&\leq \frac{N}{2} \log\left(1\! +\! \frac{1}{\bbE(\tau)} \sum_{n=1}^{\infty} \bbP(\Psi_n\! =\! 0) \bbE[(X_{1n}\! +\! X_{2n})^2|\Psi_n\! =\! 0]\right)\nn\\*
&\qquad+h_{\rmb}(\eps)\\
\label{eqn:new6}
&\leq \frac{N}{2} \log\left(1+\frac{1}{\bbE(\tau)} \sum_{n=1}^{\infty}  \bbE[(X_{1n}+X_{2n})^2]\right)\nn\\*
&\qquad+h_{\rmb}(\eps)\\
&=\frac{N}{2} \log\left(1+\frac{1}{\bbE(\tau)} \sum_{n=1}^{\infty}  \bbE[X_{1n}^2]+\bbE[X_{2n}^2]+2\bbE[X_{1n}X_{2n}]\right)\nn\\*
&\qquad+h_{\rmb}(\eps)\\
&\leq \frac{N}{2} \log\left(1+\frac{P_1 \bbE(\tau)+P_2 \bbE(\tau)}{\bbE(\tau)} \right)+h_{\rmb}(\eps)\label{eqn:setineq_end}
%&= \frac{N}{2} \log\left(1+P_1+P_2 \right)+O(1).
\end{align}
Here,~\eqref{eqn:new6} follows from the fact that $\bbE[(X_{1n}+X_{2n})^2]\ge \bbP(\Psi_n=0) \bbE[(X_{1n}+X_{2n})^2|\Psi_n=0]$.
%\begin{align}
%\bbE[(X_{1n}+X_{2n})^2]&=\bbP(\Psi_n=0) \bbE[(X_{1n}+X_{2n})^2|\Psi_n=0]+ \bbP(\Psi_n=1) \bbE[(X_{1n}+X_{2n})^2|\Psi_n=1]\\
%&\geq \bbP(\Psi_n=0) \bbE[(X_{1n}+X_{2n})^2|\Psi_n=0],
%\end{align}
and~\eqref{eqn:setineq_end} follows from the power constraints of the stop-feedback code and the fact that $X_{1n}=f_n^{(1)}(W_1)$ is independent of $X_{2n}=f_n^{(2)}(W_2)$.
%\end{IEEEproof}

Similarly, we have from~\eqref{eq171n} and~\eqref{eq128new} that
\begin{align}
&(1-\eps)\log M_1  \nn\\*
&\leq \frac{1}{2} \bbE(\tau)  \log\left(1+\sum_{n=1}^{\infty} \frac{\bbP(\Psi_n=0)}{\bbE(\tau)} \bbE[X^2_{1n}|\Psi_n=0]\right)\nn\\*
&\qquad+h_{\rmb}(\eps)\\
&\leq \frac{N}{2} \log\left(1+\frac{1}{\bbE(\tau)} \sum_{n=1}^{\infty} \bbP(\Psi_n=0) \bbE[X^2_{1n}|\Psi_n=0]\right)\nn\\*
&\qquad+h_{\rmb}(\eps)\\
\label{eqn:new8}
&\leq \frac{N}{2} \log\left(1+\frac{1}{\bbE(\tau)} \sum_{n=1}^{\infty}  \bbE[X_{1n}^2]\right)+h_{\rmb}(\eps)\\
\label{eqn:new9}
&\leq \frac{N}{2} \log\left(1+\frac{P_1 \bbE(\tau)}{\bbE(\tau)} \right)+h_{\rmb}(\eps)
%\\
%&=\frac{N}{2} \log\left(1+P_1  \right)+O(1).
\end{align}
%Here, inequalities~\eqref{eqn:new8} and~\eqref{eqn:new9} follow from the same reasoning as those in the  previous set of inequalities in \eqref{eqn:setineq1}--\eqref{eqn:setineq_end}. 
% follows from the fact that
%\begin{align}
%\bbE[X_{1n}^2]&=\bbP(\Psi_n=0) \bbE[X_{1n}^2|\Psi_n=0]+ \bbP(\Psi_n=1) \bbE[X_{1n}^2|\Psi_n=1]\\
%&\geq \bbP(\Psi_n=0) \bbE[X_{1n}^2|\Psi_n=0],
%\end{align}
%(b) follows from the power constraints of the stop-feedback code, and (c) follows from Taylor approximation.

For the case $|\calU_1| \ge 1,|\calU_2|\ge 1$, with the above arguments and $\calF_n=\sigma(U_1,U_2,\hat{Y}^n)$, the following expressions hold almost surely:
\begin{align}
\label{eq213}
&(1-\bbP[(\hat{W}_1,\hat{W}_2)\neq (W_1,W_2)|U_1,U_2])\log M_1 M_2 \nn \\*
&\leq \frac{1}{2}\log(1+P_1+P_2)\nn \\* 
&\qquad +h_{\rmb}( \bbP[(\hat{W}_1,\hat{W}_2)\neq (W_1,W_2)|U_1,U_2)]),\\
&(1-\bbP[(\hat{W}_1,\hat{W}_2)\neq (W_1,W_2)|U_1,U_2])\log M_j \nn\\*
&\leq\!  \frac{1}{2}\log(1\! +\! P_j)\! +\! h_{\rmb}( \bbP[(\hat{W}_1,\hat{W}_2)\! \neq\!  (W_1,W_2)|U_1,U_2)]),\label{eq216}
%&(1-\bbP[(\hat{W}_1,\hat{W}_2)\neq (W_1,W_2)|U_1,U_2])\log M_2 \nn\\
%\label{eq216}
%&\leq \sum_{n=1}^{\infty} \frac{1}{2}\log(1+P_2)+h_{\rmb}((\hat{W}_1,\hat{W}_2)\neq (W_1,W_2)|U_1,U_2).
\end{align}
where $j=1,2.$
By taking the expectation with respect to $(U_1, U_2)$ on both sides of~\eqref{eq213}--\eqref{eq216} and applying   Jensen's inequality  for  the binary entropy terms, we obtain~\eqref{eqn:sf_up1}--\eqref{eqn:sf_up3}. This concludes the converse proof of Theorem \ref{stop_mac}.
\end{IEEEproof}
\subsection{Converse Proof for Theorem~\ref{vlft_mac}}\label{sec:conv_v}
\begin{lemma} \label{lem:conv_mac_vlft}
Given a Gaussian MAC $\bbP(y|x_1,x_2)$, for any $0\leq \eps\leq 1-\max\{\frac{1}{M_1},\frac{1}{M_2}\}$, any $(M_1,M_2,N,P_1,P_2,\eps)$ VLFT code for any $N\in\bbN$  satisfies  \eqref{eqn:vlft_up1} and \eqref{eqn:vlft_up3} for some $\rho \in [0,1]$.
%\begin{align}
%\label{conv1_vlft}
%0\leq \log M_j &\leq \frac{N\rvC(P_j(1-\rho^2))+(N+1)h_{\rmb}\left(\frac{1}{N+1}\right)+h_{\rmb}(\eps)}{1-\eps}, \quad j=1,2 \\*
%%&\leq \frac{N\rvC(P_j(1-\rho^2))+\log(N+1)+h_{\rmb}(\eps)+1}{1-\eps}, \\
%\label{conv3_vlft}
%\log M_1M_2 & \leq \frac{N\rvC(P_1+P_2+2\rho \sqrt{P_1P_2})+(N+1)h_{\rmb}\left(\frac{1}{N+1}\right)+h_{\rmb}(\eps)}{1-\eps}.
%%&\leq \frac{N\rvC(P_1+P_2+2\rho \sqrt{P_1P_2})+\log(N+1)+h_{\rmb}(\eps)+1}{1-\eps}.
%\end{align}
\end{lemma}
\begin{IEEEproof}
Similarly to the  converse proof for Gaussian MAC with a stop-feedback code, we first consider the case in which  $|\calU_1|=|\calU_2|=1$. Since the receiver decides on the transmitted messages based only on $Y^{\tau}$ and $(W_1,W_2)$ (not dependent  on the channel outputs that are received after time $\tau$), as in~\cite{Yury2011}, we can convert any given code $(f_n^{(1)},f_n^{(2)},g_n,\tau)$ to an equivalent code $(\hat{f}_n^{(1)},\hat{f}_n^{(2)},\hat{g}_n,\tau)$ to remove the dependence of $\tau$ on $(W_1,W_2)$. To do so, we  append a special symbol $\rvT \notin \bbR$ to the input and output alphabets to form the common input-output alphabet $\bbR\cup\{\rvT\}$.
% and then assign:
%\begin{align}
%\label{eq125a}
%\bbP_{\hat{Y}|\hat{X}_1,\hat{X}_2}(\hat{y}|\hat{x}_1,\hat{x}_2)=\begin{cases} \bbP(\hat{y}|\hat{x}_1,\hat{x}_2),& \hat{x}_1 \neq T, \hat{x}_2\neq T\\1\{\hat{y}=T\},&\hat{x}_1=\hat{x}_2=T \end{cases}
%\end{align}
%and
%\begin{align}
%\hat{X}_{jn}=\hat{f}_n^{(j)}(W_j):=\begin{cases} f^{(j)}_n(W_j),&n\leq \tau\\T, &n>\tau \end{cases}, \quad j=1,2. \label{eqn:Xhat_jn}
%\end{align}
%Define
%\begin{align}
%\hat{\tau}=\tau+1=\inf\{n: \hat{Y}_n=T\}
%\end{align}
%and
%\begin{align}
%\label{eq129a}
%\hat{g}_n(\hat{Y}^n):=\begin{cases} g_n(\hat{Y}^n),&n < \hat{\tau}\\g_n(Y^{\hat{\tau}-1}),&n\geq\hat{\tau} \end{cases}.
%\end{align}
We also set $
\hat{\tau} =\tau+1=\inf\{n: \hat{Y}_n=\rvT\}$ 
%\hat{g}_n(\hat{Y}^n) &:=\begin{cases} g_n(\hat{Y}^n),&n < \hat{\tau}\\g_n(Y^{\hat{\tau}-1}),&n\geq\hat{\tau} \end{cases}.\label{eq129b}
and %\textcolor{red}{($\Psi_n$ conflicts with $T$. Change $T$ throughout to $\rvT$)}
\begin{equation}
\Psi_n:=1\{\hat{\tau}\leq n\}\in \sigma(\hat{Y}^n), \label{eqn:Psi_VLFT}
\end{equation}
which is  slightly different from the stop-feedback case (cf.\ \eqref{eqn:Psi}).

Using the same approach as the proof of converse for the Gaussian MAC with a stop-feedback code in Section \ref{sec:conv_s}, we obtain from the bounds in    Appendix \ref{app:mac} %(but retaining the term $H(\Psi_n|\hatY^{n-1})$ which does not vanish under the  current VLFT setting)  
that
\begin{align}
I(W_1,W_2;\hat{Y}_n|\hat{Y}^{n-1})&\leq H(\Psi_n|\hat{Y}^{n-1}) \nn\\*
&\hspace{-1.2in}+\frac{1}{2}\bbP(\Psi_n=0)\log(1+\bbE[(X_{1n}+X_{2n})^2|\Psi_n=0]),\\
\label{eq519n}
I(W_1;\hat{Y}_n|\hat{Y}^{n-1}W_2)&\leq H(\Psi_n|\hat{Y}^{n-1})\nn\\*
&\hspace{-1.2in}+\bbP(\Psi_n=0) I(X_{1n};Y_n|\Psi_n=0,Y^{n-1}, X_{2n}, W_2 ),\\
I(W_2;\hat{Y}_n|\hat{Y}^{n-1}W_1)&\leq H(\Psi_n|\hat{Y}^{n-1})\nn\\*
&\hspace{-1.2in}+ \bbP(\Psi_n=0) I(X_{2n};Y_n|\Psi_n=0,Y^{n-1}, X_{1n}, W_1).
\end{align} 
Note that $\{\hat{\tau}\le n-1\}$ for the stop-feedback case (cf.\ Lemma \ref{lem:convThm3}) is equivalent to $\{\hat{\tau}\le n\}$ for the VLFT case we consider here. Also compare \eqref{eqn:Psi} to \eqref{eqn:Psi_VLFT}.  Observe that
\begin{align}
&I(X_{1n};Y_n|\Psi_n=0,Y^{n-1}, X_{2n}, W_2 )\nn\\*
%&=h(Y_n|\Psi_n=0,Y^{n-1}, X_{2n}, W_2)-h(Y_n|\Psi_n=0,Y^{n-1}, X_{1n},X_{2n}, W_2)\\
%&=h(Y_n|\Psi_n=0,Y^{n-1}, X_{2n}, W_2)-h(X_{1n}+X_{2n}+Z_n|\Psi_n=0,Y^{n-1}, X_{1n},X_{2n}, W_2)\\
%&=h(Y_n|\Psi_n=0,Y^{n-1}, X_{2n}, W_2)-h(Z_n|\Psi_n=0,Y^{n-1}, X_{1n},X_{2n}, W_2)\\*
%&=h(Y_n|\Psi_n=0,Y^{n-1}, X_{2n}, W_2)-h(Z_n)\\
%\label{eq525n}
%&\leq h(Y_n|\Psi_n=0,X_{2n})-\frac{1}{2}\log(2\pi e)\\
%&=h(X_{1n}+X_{2n}+Z_n|\Psi_n=0,X_{2n})-\frac{1}{2}\log(2\pi e)\\*
&\le h(X_{1n}+Z_n|\Psi_n=0,X_{2n})-\frac{1}{2}\log(2\pi e) .\label{eqn:shortened}
\end{align}
From here on, we essentially mimic Ozarow's weak converse proof for the Gaussian MAC with fixed-length feedback~\cite{Ozarow} but with some changes in the parameter settings. First define
\begin{align}
\sigma_{jn}^2&:= \var[X_{jn} |\Psi_n=0] ,\quad j=1,2\\
\lambda_n&:=\cov[X_{1n},X_{2n}|\Psi_n=0].
\end{align}
Using the same approach as in~\cite{Ozarow}, we can show that
\begin{align}
\label{eq529n}
&h(X_{1n}+Z_n|\Psi_n=0,X_{2n})\nn\\*
&\leq \frac{1}{2}\log\left[2\pi e \sigma_{1n}^2 \left(1- \frac{\lambda_n^2}{\sigma_{1n}^2 \sigma_{2n}^2}\right)+2\pi e\right].
\end{align}
Therefore, we obtain% from~\eqref{eq519n},~\eqref{eq525n}, and~\eqref{eq529n} we obtain
\begin{align}
I(W_1,W_2;\hat{Y}_n|\hat{Y}^{n-1})&\leq H(\Psi_n|\hatY^{n-1}) \nn\\*
&\hspace{-1in}+\frac{1}{2} \bbP(\Psi_n=0)\log \left[1+\sigma_{1n}^2+\sigma_{2n}^2+2\lambda_n\right],\\
I(W_1;\hat{Y}_n|\hat{Y}^{n-1}W_2) &\leq H(\Psi_n|\hatY^{n-1})\nn\\*
&\hspace{-1in}+\frac{1}{2}\bbP(\Psi_n=0)\log\left[1+ \sigma_{1n}^2 \left(1- \frac{\lambda_n^2}{\sigma_{1n}^2 \sigma_{2n}^2}\right)\right],\\
I(W_2;\hat{Y}_n|\hat{Y}^{n-1}W_1)&\leq H(\Psi_n|\hatY^{n-1})\nn\\*
&\hspace{-1in}+\frac{1}{2}\bbP(\Psi_n=0)\log\left[1+ \sigma_{2n}^2 \left(1- \frac{\lambda_n^2}{\sigma_{1n}^2 \sigma_{2n}^2}\right)\right].
\end{align}
It follows from~\eqref{eq418newl},~\eqref{eq171n}, and~\eqref{eq173n} and the above considerations  that
\begin{align}
\label{eq170m}
(1-\eps)\log M_1M_2 &\leq \sum_{n=1}^{\infty} H(\Psi_n|\hat{Y}^{n-1}) \nn\\*
&\hspace{-1in} +\sum_{n=1}^{\infty} \frac{1}{2} \bbP(\Psi_n=0)\log \left[1+\sigma_{1n}^2+\sigma_{2n}^2+2\lambda_n\right] +h_{\rmb}(\eps),\\
\label{eq171m}
(1-\eps)\log M_1 &\leq \sum_{n=1}^{\infty} H(\Psi_n|\hat{Y}^{n-1})\nn\\*
&\hspace{-1in}+\sum_{n=1}^{\infty} \frac{1}{2}\bbP(\Psi_n=0)\log\left[1+ \sigma_{1n}^2 \left(1- \frac{\lambda_n^2}{\sigma_{1n}^2 \sigma_{2n}^2}\right)\right]+h_{\rmb}(\eps),\\
\label{eq173m}
(1-\eps)\log M_2 &\leq \sum_{n=1}^{\infty} H(\Psi_n|\hat{Y}^{n-1}) \nn\\*
& \hspace{-1in}+\sum_{n=1}^{\infty} \frac{1}{2}\bbP(\Psi_n=0)\log\left[1+ \sigma_{2n}^2 \left(1- \frac{\lambda_n^2}{\sigma_{1n}^2 \sigma_{2n}^2}\right)\right] +h_{\rmb}(\eps).
\end{align}
Note that by~\cite[Eqn.~(90)]{Yury2011}, we have
\begin{align}
\sum_{n=1}^{\infty} H(\Psi_n|\hat{Y}^{n-1})&=H(\tau) \le (N+1)h_{\rmb}\left(\frac{1}{N+1}\right)\\
%&\leq (N+1)h_{\rmb}\left(\frac{1}{N+1}\right)\\
&\leq \log(N+1)+1.
\end{align}
Moreover, since we have
\begin{align}
\sum_{n=1}^{\infty} \bbP(\Psi_n=0)&=\sum_{n=1}^{\infty} \bbP(\hat{\tau}> n)\\
&=\sum_{n=1}^{\infty} \bbP(\tau \geq n)\\
%&=\sum_{n=1}^{\infty} \bbP(\tau \geq n)\\*
&=\bbE(\tau),
\end{align}
it follows that $\{{\bbP(\Psi_n=0)}/{\bbE(\tau)}\}_{n=1}^\infty$ is a valid probability distribution. 
%\begin{align}
%\label{eq543key}
%\sum_{n=1}^{\infty} \frac{\bbP(\Psi_n=0)}{\bbE(\tau)}=1.
%\end{align}
As in Ozarow's  weak converse proof for the Gaussian MAC with fixed-length feedback~\cite{Ozarow}, the right-hand-sides of~\eqref{eq170m},~\eqref{eq171m}, and~\eqref{eq173m} can be readily shown to be  jointly concave in $(\sigma_{1n}^2, \sigma_{2n}^2,\lambda_n)$. Thus, we can use   Jensen's inequality to upper bound them.

More specifically, we set
\begin{align}
\label{eq441newq}
G_j^2&:=\sum_{n=1}^{\infty}\frac{\bbP(\Psi_n=0)}{\bbE(\tau)} \sigma_{jn}^2,\quad j=1,2\\
\label{eq442newq}
\rho&:=\frac{1}{G_1G_2}\sum_{n=1}^{\infty} \frac{\bbP(\Psi_n=0)}{\bbE(\tau)} \lambda_n. 
\end{align}
We can bound $G_1$ as follows:
\begin{align}
G_1^2&=\sum_{n=1}^{\infty}\frac{\bbP(\Psi_n=0)}{\bbE(\tau)} \sigma_{1n}^2,\\
&\leq \sum_{n=1}^{\infty}\frac{\bbP(\Psi_n=0)}{\bbE(\tau)}\bbE(X_{1n}^2|\Psi_n=0)\\
%&\leq \sum_{n=1}^{\infty}\left[\frac{\bbP(\Psi_n=0)}{\bbE(\tau)}\bbE(X_{1n}^2|\Psi_n=0)+\frac{\bbP(\Psi_n=1)}{\bbE(\tau)}\bbE(X_{1n}^2|\Psi_n=1)\right]\\
&\le\sum_{n=1}^{\infty} \frac{\bbE[X_{1n}^2]}{\bbE(\tau)}
\leq P_1.\label{eq550key}
\end{align}
The last step follows from the expected power constraints in \eqref{eqn:mac_pow}.  
Similarly, we have $G_2^2 \leq P_2.$
%\begin{align}
%\label{eq552key}
%
%\end{align}
Moreover, we also have $|\lambda_n| \leq \sigma_{1n}\sigma_{2n}$ and so from  \eqref{eq442newq} and the Cauchy-Schwarz inequality, % \textcolor{red}{I don't understand how this statement is related to the following chain}
\begin{align}
|\rho|^2 &\leq \left(\sum_{n=1}^{\infty} \frac{1}{G_1G_2} \frac{\bbP(\Psi_n=0)}{\bbE(\tau)} \sigma_{1n}\sigma_{2n}\right)^2\\
&\leq \left(\sum_{n=1}^{\infty} \frac{\bbP(\Psi_n\! =\! 0)}{\bbE(\tau)}\frac{\sigma_{1n}^2}{G_1^2} \right) \left(\sum_{n=1}^{\infty} \frac{\bbP(\Psi_n\! =\! 0)}{\bbE(\tau)}\frac{\sigma_{1n}^2}{G_2^2} \right) \\
&=1.
\end{align}
By applying Jensen's inequality to \eqref{eq170m},  we obtain
\begin{align}
&(1-\eps)\log M_1M_2 \nn\\*
&\leq \! (N \! +\! 1)h_{\rmb}\left(\frac{1}{N\! +\! 1}\right)\! +\! \frac{\bbE(\tau)}{2}\log\left[1\! +\! G_1^2\! +\! G_2^2\! +\! 2\rho G_1G_2\right]\\
&\leq \! (N\! +\! 1)h_{\rmb}\left(\frac{1}{N\! +\! 1}\right)\! + \! \frac{N}{2}\log\left[1\! +\! G_1^2\! +\! G_2^2\! +\! 2\rho G_1G_2\right]\\
&\leq\! (N\! +\! 1)h_{\rmb}\left(\frac{1}{N\! +\! 1}\right)\! + \! \frac{N}{2} \log \big[1\! +\! P_1\! +\! P_2\! +\! 2|\rho| \sqrt{P_1P_2}\big].
\end{align}
Similarly,  by applying Jensen's inequality to \eqref{eq171m} and  \eqref{eq173m}, we obtain
\begin{align}
(1-\eps)\log M_j %&\leq (N+1)h_{\rmb}\left(\frac{1}{N+1}\right)+\frac{\bbE(\tau)}{2} \log\left[1+G_j^2\left(1-\frac{(G_1G_2\rho)^2}{G_1^2 G_2^2}\right)\right]\\
%&=(N+1)h_{\rmb}\left(\frac{1}{N+1}\right)+\frac{\bbE(\tau)}{2}\log\left[1+G_j^2(1-\rho^2)\right],\\
%&\leq (N+1)h_{\rmb}\left(\frac{1}{N+1}\right)+\frac{N}{2} \log\left[1+G_j^2(1-\rho^2)\right],\\
&\leq (N+1)h_{\rmb}\left(\frac{1}{N+1}\right) \nn\\*
&\quad+\frac{N}{2} \log\big[1+P_j^2(1-\rho^2)\big],
\end{align}
for $j=1,2$.
This completes the proof of Lemma \ref{lem:conv_mac_vlft} and hence, the converse proof of Theorem~\ref{vlft_mac}.
\end{IEEEproof}
\section{Conclusion and Future Work} \label{sec:conclu}
In this paper, we derived bounds on achievable rates of the Gaussian MAC  with   the use of variable-length codes with feedback and under the  non-vanishing error probability formalism. We quantified the gains of VLFT  codes over stop-feedback codes. To establish our results, we leveraged some non-standard techniques to deal with the continuous nature of the channel and also to control the overshoot of the barrier (or threshold) of some relevant random walks. 

In the future, it would be a fruitful endeavor to improve on the second-order terms in Theorems \ref{stop_mac} and \ref{vlft_mac} as they are likely to be loose. In addition, it would be interesting to check if our newly-developed techniques for systems with variable-length feedback can be extended to other multi-terminal channel models such as the Gaussian broadcast channel.
\appendices
\section{Proof of Lemma  \ref{lem1b} } \label{app:prf} 

\begin{IEEEproof}
First, observe that
\begin{align}
&\bbE[(X_{1N}+X_{2N})^2]+\bbE[(X_{1N}-X_{2N})^2]\nn\\*
&=2(\bbE[X_{1N}^2]+\bbE[X_{2N}^2])\\
&=2[\var(X_{1N})\! +\! (\bbE X_{1N})^2 \! +\! \var(X_{2N})\! +\! (\bbE X_{2N})^2]\\
\label{eq49d}
&\leq 2[\cR_1 N+ \cS_1+o(1)+(N-A\sqrt{N}-\cD-B_1+o(1))^2 \nn\\*
&\quad + \cR_2 N+ \cS_2+o(1) + (N-A\sqrt{N}-\cD-B_2+o(1))^2].
\end{align}
Since, we have
\begin{align}
&\bbE[(X_{1N}+X_{2N})^2] \geq (\bbE[X_{1N}+X_{2N}])^2\\*
\label{eq52d}
&= (N-A\sqrt{N}-\cD-B_1+o(1)\nn\\*
&\qquad+N-A\sqrt{N}-\cD-B_2+o(1))^2.
\end{align}
It follows from~\eqref{eq49d} and~\eqref{eq52d} that
\begin{align}
&\bbE[(X_{1N}-X_{2N})^2]\nn\\*
 &\leq 2[\cR_1 N+ \cS_1+o(1)+(N-A\sqrt{N}-\cD-B_1+o(1))^2 \nn\\
&\quad + \cR_2 N+ \cS_2+o(1) + (N-A\sqrt{N}-\cD-B_2+o(1))^2]\nn\\
&\quad -(N-A\sqrt{N}-\cD-B_1+o(1) \nn\\*
&\qquad+N-A\sqrt{N}-\cD-B_2+o(1))^2 \\
&= 2[\cR_1 N+ \cS_1 +o(1)+ \cR_2 N+ \cS_2+o(1)] \nn\\*
&\quad+(B_1-B_2+o(1))^2\\
&\le 2[\cR_1+\cR_2]N+2(\cS_1+\cS_2)+(B_1-B_2)^2+o(1) \\
&\le 2[\cR_1+\cR_2]N+ 2|\cS_1+\cS_2|+(B_1-B_2)^2+o(1).
\end{align} 
Therefore, we have
\begin{align}
&(\bbE|X_{1N}-X_{2N}|)^2 \leq \bbE[(X_{1N}-X_{2N})^2] \\
&\leq 2[\cR_1+\cR_2]N+ 2|\cS_1+\cS_2|+(B_1-B_2)^2+o(1).
\end{align}
By using the fact that $(a+b)^{1/2}\le a^{1/2}+b^{1/2}$ for nonnegative $a,b$, it follows that
\begin{align}
\label{eq57d}
&\bbE|X_{1N}-X_{2N}| \leq \sqrt{2(\cR_1+\cR_2)N}\nn\\*
&\qquad+\sqrt{2|\cS_1+\cS_2|+(B_1-B_2)^2}+ o(1).
\end{align}
Similarly, we have
\begin{align}
\label{eq58d}
&\bbE|X_{iN}-X_{jN}|  \leq \sqrt{2(\cR_i+\cR_j) N} \nn\\*
&\qquad+\sqrt{2|\cS_i+\cS_j|+(B_i-B_j)^2}+o(1).
\end{align} for any $(i,j) \in \{1,2,3\}\times\{1,2,3\}$.

Now, we note that
\begin{align}
\max\{X_{iN},X_{jN}\}&=\frac{1}{2}[X_{iN}+X_{jN}+|X_{iN}-X_{jN}|]
\end{align} for any $(i,j) \in \{1,2,3\}\times\{1,2,3\}$.

Therefore, we have
\begin{align}
\label{eq422supp}
&\max\{X_{1N},X_{2N},X_{3N}\}\nn\\*
&=\max\{\max\{X_{1N},X_{2N}\},X_{3N}\}\\
&=\frac{1}{2} \max\{X_{1N}+X_{2N}+|X_{1N}-X_{2N}|,2 X_{3N}\}\\
&=\frac{1}{4}\Big[X_{1N}+X_{2N}+|X_{1N}-X_{2N}|+2X_{3N} \nn\\
&\quad +|(X_{1N}+X_{2N}+|X_{1N}-X_{2N}|)-2X_{3N}|\Big]\\
&=\frac{1}{4}\Big[(X_{1N}+X_{2N}+2X_{3N})+|X_{1N}-X_{2N}|\nn\\
&\quad +|(X_{1N}\! -\! X_{3N})\! +\! (X_{2N}\! -\! X_{3N})\! +\!  |X_{1N}\! -\! X_{2N}| \Big] \\
&\leq \frac{1}{4}\Big[(X_{1N}+X_{2N}+2X_{3N})+2|X_{1N}-X_{2N}|\nn\\*
&\quad +|X_{1N}-X_{3N}|+|X_{2N}-X_{3N}|\Big].
\end{align}
It follows that
\begin{align}
&\bbE[\max\{X_{1N},X_{2N},X_{3N}\}]\nn\\*
&\leq \frac{1}{4}\bbE[X_{1N}+X_{2N}+2X_{3N}]\nn\\*
&\quad +\frac{1}{4} \bbE\left(2|X_{1N}-X_{2N}|+|X_{1N}-X_{3N}|+|X_{2N}-X_{3N}|\right)\\
&=\frac{1}{4} \Big(\bbE[X_{1N}]+\bbE[X_{2N}]+2\bbE[X_{3N}]\Big)\nn\\
&\quad + \frac{1}{2}\Big(\bbE|X_{1N}-X_{2N}|\Big) \nn\\*
&\quad+\frac{1}{4}\Big(\bbE|X_{1N}-X_{3N}|+\bbE|X_{2N}-X_{3N}|\Big)\\
&\leq N-A\sqrt{N}-\cD-\frac{1}{4}\left(B_1+B_2+2B_3\right)+o(1)\nn\\*
&\quad + \frac{1}{2} \Big[\sqrt{2(\cR_1\! +\! \cR_2)}\sqrt{N}\! +\! \sqrt{2|\cS_1\! +\! \cS_2|\! +\! (B_1\! -\! B_2)^2}\! +\!  o(1)\Big] \nn\\*
&\quad +\frac{1}{4} \Big[\sqrt{2(\cR_1\! +\! \cR_3)}\sqrt{N}\! +\! \sqrt{2|\cS_1\! +\! \cS_3|\! +\! (B_1\! -\! B_3)^2}+ o(1) \nn\\*
&\quad +\sqrt{2(\cR_2\! +\! \cR_3)}\sqrt{N}\! +\!  \sqrt{2|\cS_2\! +\! \cS_3|\! +\! (B_2\! -\! B_3)^2}\! +\!  o(1)\Big]\\
&=N-\sqrt{N}\Big[A-\frac{1}{2} \sqrt{2(\cR_1+\cR_2)} -\frac{1}{4}(\sqrt{2(\cR_1+\cR_3)}\nn\\*
&\quad+ \sqrt{2(\cR_2+\cR_3)})\Big]-\cD-\frac{1}{4}\left(B_1+B_2+2B_3\right)\nn\\*
&\quad+\frac{1}{2}\left(\sqrt{2|\cS_1+\cS_2|+(B_1-B_2)^2}\right)\nn\\*
&\quad+\frac{1}{4}\Big(\sqrt{2|\cS_1+\cS_3|+(B_1-B_3)^2}\nn\\*
&\quad+ \sqrt{2|\cS_2+\cS_3|+(B_2-B_3)^2}\Big)+o(1). \label{eqn:chooseAG}
\end{align}
Now, if we choose
\begin{align}
  A = \frac{1}{2}\sqrt{2(\cR_1 + \cR_2)} +\frac{1}{4}\left(\sqrt{2(\cR_1+ \cR_3)}\! +\!  \sqrt{2(\cR_2+ \cR_4)}\right)
\end{align}
and
\begin{align}
\cD& =-\frac{1}{4}\left(B_1\! +\! B_2\! +\! 2B_3\right)+\frac{1}{2}\left(\sqrt{2|\cS_1\! +\! \cS_2|+(B_1 \! -\! B_2)^2}\right)\nn\\
&\quad+\frac{1}{4}\Big(\sqrt{2|\cS_1+\cS_3|+(B_1-B_3)^2}\nn\\*
&\quad+ \sqrt{2|\cS_2+\cS_3|+(B_2-B_3)^2}\Big),
\end{align}   from \eqref{eqn:chooseAG}, we   have
\begin{align}
\bbE[\max\{X_{1N},X_{2N},X_{3N}\}]\leq N+o(1).
\end{align}
Notice the symmetry of  $X_{1N},X_{2N},X_{3N}$ in the expression $\max\{X_{1N},X_{2N},X_{3N}\}$. Hence, by the above approximation procedure, the smallest value  of $A$ that we can choose is given by~\eqref{para_anewnew}. The proof of  Lemma \ref{lem1b} can now be completed by choosing the order of combination $X_{1N},X_{2N}, X_{3N}$ in~\eqref{eq422supp} such that $A$ is minimized.
\end{IEEEproof}

\section{Bounds on Mutual Information Quantities for the Gaussian MAC} \label{app:mac}
%We now  prove the following lemma:
\begin{lemma}
\label{lem3}
For any stop-feedback code for the Gaussian MAC as in Definition~\ref{def1_mac} and its equivalent form with the augmented symbol $\rvT$ for the case $|\calU_1|=|\calU_2|=1$, define
$
\Psi_n:=1\{\hat{\tau} \leq n-1\}\in \sigma(\hat{Y}^{n-1})
$  (cf.\ \eqref{eqn:Psi}). 
Then the following  bounds hold:
\begin{align}
\label{eq127}
&I(W_1W_2;\hat{Y}^{\infty}) \nn\\* 
&\leq \sum_{n=1}^{\infty} \frac{1}{2}\bbP(\Psi_n=0)\log(1+\bbE[(X_{1n}+X_{2n})^2|\Psi_n=0]),\\
\label{eq128}
&I(W_1;\hat{Y}^{\infty}|W_2)\nn\\* 
&\leq \sum_{n=1}^{\infty} \frac{1}{2} \bbP(\Psi_n=0)\log(1+\bbE[X_{1n}^2|\Psi_n=0]),\\
\label{eq129}
&I(W_2;\hat{Y}^{\infty}|W_1)\nn\\* 
&\leq \sum_{n=1}^{\infty} \frac{1}{2} \bbP(\Psi_n=0)\log(1+\bbE[X_{2n}^2|\Psi_n=0]).
\end{align}
\end{lemma}
\begin{IEEEproof}
To prove~\eqref{eq127}, we observe that
\begin{align}
I(W_1W_2;\hat{Y}^{\infty})&=\sum_{n=1}^{\infty} I(W_1,W_2;\hat{Y}_n|\hat{Y}^{n-1}).
\end{align}
Consider,
\begin{align}
&I(W_1,W_2;\hat{Y}_n|\hat{Y}^{n-1})\nn\\
&= I(W_1,W_2;\hat{Y}_n,\Psi_n|\hat{Y}^{n-1})\\
&= I(W_1,W_2;\Psi_n|\hat{Y}^{n-1})+I(W_1,W_2;\hat{Y}_n|\Psi_n,\hat{Y}^{n-1})\\
\label{eqn:new11}
&\leq H(\Psi_n|\hat{Y}^{n-1})+ I(W_1,W_2;\hat{Y}_n|\Psi_n,\hat{Y}^{n-1})\\
\label{eqn:new12}
&= I(W_1,W_2;\hat{Y}_n|\Psi_n,\hat{Y}^{n-1})\\
%&= \bbP(\Psi_n=0) I(W_1,W_2;\hat{Y}_n|\Psi_n=0,\hat{Y}^{n-1}) + \bbP(\Psi_n=1) I(W_1,W_2;\hat{Y}_n|\Psi_n=1,\hat{Y}^{n-1})\\
\label{eqn:new13}
&= \bbP(\Psi_n=0) I(W_1,W_2;\hat{Y}_n|\Psi_n=0,\hat{Y}^{n-1})\\
&\leq \bbP(\Psi_n=0) I(\hat{X}_{1n},\hat{X}_{2n};\hat{Y}_n|\Psi_n=0,\hat{Y}^{n-1})\\
\label{eq473key}
&= \bbP(\Psi_n=0) I(X_{1n},X_{2n};Y_n|\Psi_n=0,Y^{n-1})\\
&= \bbP(\Psi_n=0) [h(Y_n|\Psi_n=0,Y^{n-1}) \nn\\* 
&\qquad-h(Y_n|X_{1n},X_{2n},\Psi_n=0,Y^{n-1})]\\
&\leq \bbP(\Psi_n=0) [h(Y_n|\Psi_n=0)\nn\\* 
&\qquad-h(Y_n|X_{1n},X_{2n},\Psi_n=0,Y^{n-1})]\\
%&= \bbP(\Psi_n=0) [h(Y_n|\Psi_n=0)-h(X_{1n}+X_{2n}+Z_n|X_{1n},X_{2n},\Psi_n=0, Y^{n-1})]\\
&= \bbP(\Psi_n=0) [h(Y_n|\Psi_n=0) \nn\\* 
&\qquad-h(Z_n|X_{1n},X_{2n},\Psi_n=0, Y^{n-1})]\\
\label{eqn:new14}
&\leq \bbP(\Psi_n=0) [h(Y_n|\Psi_n=0)-h(Z_n)]\\
\label{eqn:new15}
&\leq  \bbP(\Psi_n\!=\! 0) \left[\frac{1}{2}\!\log[2\pi e \bbE(Y_n^2|\Psi_n\!=\! 0)]\!-\!\frac{1}{2}\log [2\pi e]\right]\\
%&=\frac{1}{2}\bbP(\Psi_n=0) \log[\bbE(Y_n^2|\Psi_n=0)]\\
&=\frac{1}{2}\bbP(\Psi_n=0) \log[\bbE(X_{1n}+X_{2n}+Z_n)^2|\Psi_n=0]\\
&=\frac{1}{2}\bbP(\Psi_n=0) \log[\bbE((X_{1n}+X_{2n})^2|\Psi_n=0) \nn\\*
&\qquad+\bbE(X_{1n}Z_n|\Psi_n=0)+\bbE(X_{2n}Z_n|\Psi_n=0)\nn\\* 
&\qquad+\bbE(Z_n^2|\Psi_n=0)]\\
\label{eq473key_a}
&=\frac{1}{2}\bbP(\Psi_n=0)\log[1+\bbE((X_{1n}+X_{2n})^2|\Psi_n=0)], %\label{eqn:499}
\end{align} where~\eqref{eqn:new11} follows from the fact that $\Psi_n$ is a binary random variable,~\eqref{eqn:new12} follows from the fact that $\Psi_n \in \sigma(\hat{Y}^{n-1})$,~\eqref{eqn:new13} follows from the fact that given $\Psi_n=1$ or $n \geq \hat{\tau}+1$ we always have $\hat{Y}_n=\rvT$,~\eqref{eq473key} follows from the fact that given $\Psi_n=0$ or $\tau \geq n$ we have $\hat{X}_{1n}=X_{1n}, \hat{X}_{2n}=X_{2n}$, and $\hat{Y}_n=Y_n$,~\eqref{eqn:new14} follows from the fact that $\Psi_n=1\{\hat{\tau} \leq n-1\}=1\{\tau \leq n-1\}$ is a function of $\sigma(Y^{n-1}), X_{1n}=f_n^{(1)}(W_1), X_{2n}=f_n^{(1)}(W_2)$ and $Z_n$ is independent of $(Y^{n-1},W_1,W_2)$,~\eqref{eqn:new15} follows from the maximal differential entropy formula,~\eqref{eq473key_a} follows from the facts that $\Psi_n$ is a function of $Y^{n-1}$ and $Z_n$ is independent of $(X_{1n},X_{2n},Y^{n-1})$. 

It follows that
\begin{align}
&I(W_1W_2;\hat{Y}^{\infty})\nn\\*
&\leq \sum_{n=1}^{\infty} \frac{1}{2}\bbP(\Psi_n=0)\log[1+\bbE(X_{1n}+X_{2n})^2|\Psi_n=0].
\end{align}
The other inequalities can be shown in a completely analogous manner. 
%Now, we have
%\begin{align}
%I(W_1;\hat{Y}^{\infty}|W_2)=\sum_{n=1}^{\infty} I(W_1;\hat{Y}_n|W_2\hat{Y}^{n-1}).
%\end{align}
%Using similar arguments, we have
%\begin{align}
%I(W_1;\hat{Y}_n|\hat{Y}^{n-1}W_2)
%\label{eq492key}
% &\leq I(W_1;\hat{Y}_n\Psi_n|\hat{Y}^{n-1},W_2)\\
%&\leq \frac{1}{2} \bbP(\Psi_n=0)\log(1+\bbE[X_{1n}^2|\Psi_n=0]).
%\end{align} It follows that
%\begin{align}
%I(W_1;\hat{Y}^{\infty}|W_2)\leq \sum_{n=1}^{\infty} \frac{1}{2} \bbP(\Psi_n=0)\log(1+\bbE[X_{1n}^2|\Psi_n=0]).
%\end{align}
%Similarly, we also have
%\begin{align}
%I(W_2;\hat{Y}^{\infty}|W_1)\leq \sum_{n=1}^{\infty} \frac{1}{2} \bbP(\Psi_n=0)\log(1+\bbE[X_{2n}^2|\Psi_n=0]).
%\end{align}
%This completes the proof of Lemma \ref{lem3}.
\end{IEEEproof} 
\subsection*{Acknowledgements}
The authors would like to sincerely thank Dr.\ Baris Nakibo\u{g}lu for bringing our attention to references \cite{Burnashev80} and~\cite{Nak08}.  The authors are also extremely grateful to  the associate editor Prof.\ Aslan Tchamkerten and the four anonymous reviewers for their excellent and detailed comments that helped to streamline the presentation of the paper.
\bibliographystyle{unsrt}
\bibliography{isitbib}

\begin{thebibliography}{10}

\bibitem{Sha56}
C.~E. Shannon.
\newblock The zero error capacity of a noisy channel.
\newblock {\em IRE Trans. on Inform. Th.}, 2(3):8--19, 1956.

\bibitem{SK66}
J.~Schalkwijk and T.~Kailath.
\newblock A coding scheme for additive noise channels with feedback--{Part I}:
  No bandwith constraint.
\newblock {\em IEEE Trans. on Inform. Th.}, 12(2):172--182, 1966.

\bibitem{Burn14}
M.~Burnashev and H.~Yamamoto.
\newblock On using feedback in a {Gaussian} channel.
\newblock {\em Problems of Information Transmission}, 50(3):19--34, 2014.

\bibitem{Ozarow}
L.~H. Ozarow.
\newblock The capacity of the white {Gaussian multiple} access channel with
  feedback.
\newblock {\em IEEE Trans. on Inform. Th.}, 30(4):623--629, 1984.

\bibitem{ShayevitzF}
O.~Shayevitz and M.~Feder.
\newblock Optimal feedback communication via posterior matching.
\newblock {\em IEEE Trans. on Inform. Th.}, 57(3):1186--1222, 2011.

\bibitem{TruongFongTan17}
L.~V. Truong, S.~L. Fong, and V.~Y.~F. Tan.
\newblock On {Gaussian} channels with feedback under expected power constraints
  and with non-vanishing error probabilities.
\newblock {\em IEEE Trans. on Inform. Th.}, 63(3):1746--1765, Mar 2017.

\bibitem{CoverPombra1989}
T.~Cover and S.~Pombra.
\newblock Gaussian feedback capacity.
\newblock {\em IEEE Trans. on Inform. Th.}, 35(1):37--43, 1989.

\bibitem{YHK2010}
Y.~H. Kim.
\newblock Feedback capacity of stationary {Gaussian} channels.
\newblock {\em IEEE Trans. on Inform. Th.}, 56(1):57--85, 2010.

\bibitem{HaimPermuter2008}
H.~Permuter, P.~Cuff, B.~{Van Roy}, and T.~Weissman.
\newblock Capacity of the trapdoor channel with feedback.
\newblock {\em IEEE Trans. on Inform. Th.}, 54(7):3150 -- 3165, 2008.

\bibitem{HaimPermuter2014}
O.~Elischo and H.~Permuter.
\newblock Capacity and coding for the {Ising} channel with feedback.
\newblock {\em IEEE Trans. on Inform. Th.}, 60(9):5138--5149, 2014.

\bibitem{AW14}
Y.~Altu\u{g} and A.~B. Wagner.
\newblock Feedback can improve the second-order coding performance in discrete
  memoryless channels.
\newblock In {\em Proc. of Intl. Symp. on Inform. Th.}, pages 2361--2365,
  Honolulu, HI, 2014.

\bibitem{Burnashev1976}
M.~V. Burnashev.
\newblock Data transmission over a discrete channel with feedback. {Random}
  transmission time.
\newblock {\em Problems of Information Transmission}, 12(4):10--30, 1976.

\bibitem{YamamotoItoh1979}
H.~Yamamoto and K.~Itoh.
\newblock Asymptotic performance of a modified {Schalkwijk-Barron} scheme for
  channels with noiseless feedback.
\newblock {\em IEEE Trans. on Inform. Th.}, 25(6):729--733, 1979.

\bibitem{TanBook}
V.~Y.~F. Tan.
\newblock Asymptotic estimates in information theory with non-vanishing error
  probabilities.
\newblock {\em {Foundations and Trends in Communications and Information
  Theory}}, 11(1--2):1--184, Sep 2014.

\bibitem{Yury2011}
Y.~Polyanskiy, H.~V. Poor, and S.~Verd\'u.
\newblock Feedback in the non-asymptotic regime.
\newblock {\em IEEE Trans. on Inform. Th.}, 57(8):4903--4925, 2011.

\bibitem{Trillingsgaard:2014}
K.~F. Trillingsgaard and P.~Popovski.
\newblock Variable-length coding for short packets over a multiple access
  channel with feedback.
\newblock In {\em Proc. 11th Intl. Symp. on Wireless Communications Systems},
  pages 796--800, Barcelona, Spain, 2014.

\bibitem{TK14}
V.~Y.~F. Tan and O.~Kosut.
\newblock On the dispersions of three network information theory problems.
\newblock {\em IEEE Trans. on Inform. Th.}, 60(2):881--903, 2014.

\bibitem{Mol12}
E.~MolavianJazi and J.~N. Laneman.
\newblock A second-order achievable rate region for {Gaussian} multi-access
  channels via a central limit theorem for functions.
\newblock {\em IEEE Trans. on Inform. Th.}, 61(12):6719--6733, Dec 2015.

\bibitem{Devassy2016}
R.~Devassy, G.~Durisi, B.~Lindqvist, W.~Yang, and M.~Dalai.
\newblock Nonasymptotic coding-rate bounds for binary erasure channels with
  feedback.
\newblock {\em Information Theory (cs.IT)}, 2016.
\newblock {\tt arXiv:1607.06837 [cs.IT]}.

\bibitem{Pol13b}
Y.~Polyanskiy.
\newblock Dispersion of compound channels.
\newblock In {\em Proc. of Allerton Conference}, Monticello, IL, 2013.

\bibitem{Trillingsgaard15}
K.~F. Trillingsgaard, W.~Yang, G.~Durisi, and P.~Popovski.
\newblock Broadcasting a common message with variable-length stop-feedback
  codes.
\newblock In {\em Proc. of Intl. Symp. on Inform. Th.}, pages 2505--2509, Hong
  Kong, China, 2015.

\bibitem{Trillingsgaard2016}
K.~F. Trillingsgaard, W.~Yang, G.~Durisi, and P.~Popovski.
\newblock Variable-length coding with stop-feedback for the common-message
  broadcast channel in the nonasymptotic regime.
\newblock {\em Information Theory (cs.IT)}, 2016.
\newblock {\tt arXiv:1607.03519 [cs.IT]}.

\bibitem{Burnashev80}
M.~V. Burnashev.
\newblock Sequential discrimination of hypotheses with control of observations.
\newblock {\em Math. USSR Izv.}, 15(3):419--440, 1980.

\bibitem{Nak08}
B.~Nakibo\u{g}lu and R.~G. Gallager.
\newblock Error exponents for variable-length block codes with feedback and
  cost constraints.
\newblock {\em IEEE Trans. on Inform. Th.}, 54(3):945--963, 2008.

\bibitem{Williams}
D.~Williams.
\newblock {\em Probabilities with Martingales}.
\newblock Cambridge Univ. Press, 1991.

\bibitem{Gut1974}
A.~Gut.
\newblock On the moments and limit distributions of some first passage times.
\newblock {\em The Annals of Probability}, 2(2):277--308, 1974.

\bibitem{LaiSiegmund1979}
T.~L. Lai and D.~Siegmund.
\newblock A nonlinear renewal theory with applications to sequential analysis
  {II}.
\newblock {\em The Annals of Statistics}, 7(1):60--76, 1979.

\bibitem{elgamal}
A.~{El~Gamal} and Y.-H. Kim.
\newblock {\em Network Information Theory}.
\newblock Cambridge University Press, Cambridge, U.K., 2012.

\bibitem{FongTan16}
S.~L. Fong and V.~Y.~F. Tan.
\newblock A proof of the strong converse theorem for {Gaussian} multiple access
  channels.
\newblock {\em IEEE Trans. on Inform. Th.}, 62(8):4376--4394, Aug 2016.

\bibitem{Stone1965}
C.~Stone.
\newblock On characteristic functions and renewal theory.
\newblock {\em Trans. Amer. Math. Soc.}, 120(2):327--342, 1965.

\bibitem{Grimmett}
G.~R. Grimmett and D.~R. Stirzaker.
\newblock {\em Probability and Random Processes}.
\newblock Oxford Science Publications, 2nd edition, 1992.

\bibitem{wikiwald}
Wikipedia.
\newblock Wald's equation.
\newblock
  \url{https://en.wikipedia.org/wiki/Wald}.

\bibitem{personalCommAE}
A.~Tchamkerten.
\newblock Personal communication, Apr 2017.

\bibitem{Bruss}
F.~{Thomas Bruss} and J.~B. Robertson.
\newblock {`Wald's Lemma'} for sums of order statistics of i.i.d. random
  variables.
\newblock {\em Advances in Applied Probability}, 23(3):612--623, 1991.

\bibitem{Battacharya}
R.~N. Bhattacharya and R.~R. Rao.
\newblock {\em Normal Approximation and Asymptotic Expansions}.
\newblock Wiley, New Jersey, United States, 1976.

\bibitem{Billingsley}
P.~Billingsley.
\newblock {\em Probability and Measure}.
\newblock Wiley-Interscience, 3rd edition, 1995.

\end{thebibliography}
 
\begin{IEEEbiographynophoto}{Lan V. Truong} (S'12-M'15) received the B.S.E.\ degree in Electronics
and Telecommunications from Posts and Telecommunications Institute of
Technology (PTIT), Hanoi, Vietnam in 2003. After several years of working
as an operation and maintenance engineer (O\&M) at MobiFone Telecommunications
Corporation, Hanoi, Vietnam, he resumed his graduate studies at
School of Electrical and Computer Engineering (ECE), Purdue University,
West Lafayette, IN, United States and got the M.S.E.\ degree in 2011. From
2013 to June 2015, he was an academic lecturer at Department of Information
Technology Specialization (ITS), FPT University, Hanoi, Vietnam. Since
August 2015, he has been working as a Ph.D.\ student at Department of
Electrical \& Computer Engineering (ECE), National University of Singapore
(NUS), Singapore. His research interests include information theory, coding theory, and communications.
\end{IEEEbiographynophoto}

\begin{IEEEbiographynophoto}{Vincent Y.\ F.\ Tan} (S'07-M'11-SM'15) was born in Singapore in 1981. He is currently an Assistant Professor in the Department of Electrical and Computer Engineering (ECE) and the Department of Mathematics at the National University of Singapore (NUS). He received the B.A.\ and M.Eng.\ degrees in Electrical and Information Sciences from Cambridge University in 2005 and the Ph.D.\ degree in Electrical Engineering and Computer Science (EECS) from the Massachusetts Institute of Technology in 2011. He was a postdoctoral researcher at the University of Wisconsin-Madison and a research scientist at the Institute for Infocomm (I$^2$R) Research, A*STAR, Singapore. His research interests include information theory and machine learning.

Dr.\ Tan received the MIT EECS Jin-Au Kong outstanding doctoral thesis prize in 2011, the NUS Young Investigator Award in 2014, the NUS Engineering Young Researcher Award in 2018, and the Singapore National Research Foundation (NRF) Fellowship (Class of 2018). He has authored a research monograph on {\em ``Asymptotic Estimates in Information Theory with Non-Vanishing Error Probabilities''} in the Foundations and Trends in Communications and Information Theory Series (NOW Publishers). He is currently an Editor of the IEEE Transactions on Communications and a Guest Editor for the IEEE Journal of Selected Topics in Signal Processing. 
\end{IEEEbiographynophoto}

\end{document}